\documentclass{aastex63}
\usepackage{apjfonts}
\usepackage{float}
\usepackage{commath}
\usepackage{amsmath}
\usepackage{url}
\usepackage{graphicx}
\usepackage{enumitem}
\DeclareSymbolFont{matha}{OML}{txmi}{m}{it}
\DeclareMathSymbol{\varv}{\mathord}{matha}{29}

\newcommand\msol{$M_{\odot}$}%
\def\simgt{\lower.5ex\hbox{$\; \buildrel > \over \sim \;$}}%
\def\simlt{\lower.5ex\hbox{$\; \buildrel < \over \sim \;$}}%
\def\ta{Type Ia}%
\def\snia{SNe Ia}%
\def\sni{SN Ia}%
\def\kspn{KSP-OT-201509b}%
\def\iaun{AT2015cx}%
\def\ni56{${\rm{^{56}Ni}}$}%
\def\mej{$M_{\rm ej}$}%
\def\eej{$E_{\rm ej}$}%
\def\vej{$\varv_{\rm ej}$}%
\def\co56{${\rm{^{56}Co}}$}%
\def\dm15{$\Delta$$M_{B,15}$}%
\def\pp{\rm Phillips parameter}%
\def\subch{\rm sub-Chandrasekhar}%
\def\rapd{\rm rapidly-declining}%
\def\sbv{$s_{BV}$}%
\def\ergs{\rm erg s$^{-1}$}%
\def\csp{\rm color stretch parameter}%
\def\tp{$t_{\rm p}$}%
\def\t0{$t_{\rm 0}$}%
\def\x1{$x_1$}%
\def\kms{$\rm km\;s^{\rm -1}$}%
\def\galgg{\textbf G}%
\def\snp{SNooPy}%
\def\chisqr{$\chi^2_{\rm R}$}%
\def\mpas{$\rm mag\;arcsec^{-2}$}%
\def\vi{\mbox{$V\!-\!I$}}%
\newcommand{\moon}[1]{{\rm{#1}}}

\accepted{Februaary 2, 2021}
\submitjournal{ApJ}

\shorttitle{Transitional Hostless Type~Ia Supernova}
\shortauthors{Moon et al.}

\begin{document}

\title{Rapidly-Declining Hostless Type Ia Supernova \kspn\ from the KMTNet Supernova Program: \\ Transitional Nature and Constraint on \ni56 Distribution and Progenitor Type}

\correspondingauthor{Dae-Sik Moon}
\email{moon@astro.utoronto.ca}

\author[0000-0003-4200-5064]{Dae-Sik Moon}
\affiliation{David A. Dunlap Department of Astronomy and Astrophysics, University of Toronto, 50 St. George Street, Toronto, ON M5S 3H4, Canada}

\author[0000-0003-3656-5268]{Yuan Qi Ni}
\affiliation{David A. Dunlap Department of Astronomy and Astrophysics, University of Toronto, 50 St. George Street, Toronto, ON M5S 3H4, Canada}

\author[0000-0001-7081-0082]{Maria R. Drout}
\affiliation{David A. Dunlap Department of Astronomy and Astrophysics, University of Toronto, 50 St. George Street, Toronto, ON M5S 3H4, Canada}
\affiliation{The Observatories of the Carnegie Institution for Science, 813 Santa Barbara St., Pasadena, CA 91101, USA}

\author[0000-0001-9541-0317]{Santiago Gonz\'alez-Gait\'an}
\affiliation{Centro de Modelamiento Matem\'atico, Universidad de Chile, Beauchef 851, Piso 7, Santiago, Chile}
\affiliation{CENTRA, Instituto Superior T\'ecnico - Universidade de Lisboa, Portugal}

\author[0000-0002-1338-490X]{Niloufar Afsariardchi}
\affiliation{David A. Dunlap Department of Astronomy and Astrophysics, University of Toronto, 50 St. George Street, Toronto, ON M5S 3H4, Canada}

\author[0000-0002-3505-3036]{Hong Soo Park}
\affiliation{Korea Astronomy and Space Science Institute, 776, Daedeokdae-ro, Yuseong-gu, Daejeon, 34055, Republic of Korea}

\author[0000-0002-6261-1531]{Youngdae Lee}
\affiliation{Korea Astronomy and Space Science Institute, 776, Daedeokdae-ro, Yuseong-gu, Daejeon, 34055, Republic of Korea}
\affiliation{Department of Astronomy and Space Science, Chungnam National University, Daejeon 34134, Republic of Korea}

\author[0000-0001-9670-1546]{Sang Chul Kim}
\affiliation{Korea Astronomy and Space Science Institute, 776, Daedeokdae-ro, Yuseong-gu, Daejeon, 34055, Republic of Korea}
\affiliation{Korea University of Science and Technology (UST), Daejeon 34113, Republic of Korea}

\author[0000-0003-4453-3776]{John Antoniadis}
\affiliation{Institute of Astrophysics, FORTH, Department of Physics, University of Crete, Voutes, University Campus, Heraklion, Greece}
\affiliation{Max-Planck-Institut f\"ur Radioastronomie, Bonn, DE}

\author[0000-0002-4292-9649]{Dong-Jin Kim}
\affiliation{Korea Astronomy and Space Science Institute, 776, Daedeokdae-ro, Yuseong-gu, Daejeon, 34055, Republic of Korea}

\author[0000-0001-7594-8072]{Yongseok Lee}
\affiliation{Korea Astronomy and Space Science Institute, 776, Daedeokdae-ro, Yuseong-gu, Daejeon, 34055, Republic of Korea}
\affiliation{School of Space Research, Kyung Hee University, Yongin 17104, Republic of Korea}

\begin{abstract}
We report the early discovery and multi-color ($BVI$) high-cadence 
light curve analyses of a \rapd\ \subch\ Type Ia supernova \kspn\ (= \iaun) 
from the KMTNet Supernova Program.
The Phillips parameter and color stretch parameter of \kspn\
are \dm15\ $\simeq$ 1.62 mag and \sbv\ $\simeq$ 0.54, respectively,
at an inferred redshift of 0.072.
These, together with other measured parameters (such as the strength of the
secondary $I$-band peak, colors and luminosity),
identify the source to be a \rapd\ Type Ia of transitional nature  
that is closer to Branch Normal than 91bg-like.
Its early light curve evolution and bolometric luminosity 
are consistent with those of homologously expanding ejecta powered by radioactive decay
and a Type Ia SN explosion with 0.32 $\pm$ 0.01 \msol\ of  synthesized \ni56\ mass, 
0.84 $\pm$ 0.12 \msol\ of ejecta mass 
and (0.61 $\pm$ 0.14) $\times$ 10$^{51}$ erg of ejecta kinetic energy.
While its $B-V$ and $V-I$ colors evolve largely synchronously 
with the changes in the $I$-band light curve 
as found in other supernovae,
we also find the presence of an
early redward evolution in \vi\ prior to --10 days since peak.
The bolometric light curve of the source is compatible with 
a stratified \ni56\ distribution extended to shallow layers 
of  the exploding progenitor.
Comparisons between the observed light curves 
and those predicted from ejecta-companion interactions 
clearly disfavor Roche Lobe-filling companion stars at large separation distances,
thus supporting a double-degenerate scenario for its origin.
The lack of any apparent host galaxy in our deep stack images reaching 
a sensitivity limit of $\sim$ 28 \mpas\
makes \kspn\ a hostless \ta\ supernova and offers
new insights into supernova host galaxy environments.
\end{abstract}


\section{Introduction}\label{sec:intro}

It is well established that Type Ia supernovae (SNe Ia) originate from 
explosive thermonuclear runaways in a C-O white dwarf \citep[WD;][]{hf60},
usually attributed to mass accretion from a non-degenerate binary companion
or a merger/collision of two WDs. 
The former is known as the single-degenerate \citep[e.g.,][]{wi73}
scenario; the latter as the double-degenerate scenario \citep[e.g.,][]{it84,web84}.
Although the overall similarities in observed light curves of \snia\
are indicative of the presence of shared underlying core mechanisms,
there exists substantial diversity in other observed properties,
including uncertainties in the inferred accretion processes, explosion mechanisms, 
nature of progenitors as well as host galaxy environment. 
According to \citet{liet01}, for example, about one third of nearby \snia\ are peculiar. 
\moon{The cause of this diversity is not well understood and under considerable debate
\citep[see][for a recent review on peculiar \snia]{tau17}.}

Among these stands out the long-lasting question about the origin of 
the group of \snia\ exhibiting rapid post-peak evolution
identified with a large decline rate in light curves. 
This group of \snia\ is called ``91bg-like" following its archetype SN 1991bg \citep{fet92,leiet93}, 
while the main population are ``Branch Normal."
When viewed on the Phillips relation \citep{phi93,phiet99},
which compares the peak luminosity and the decline rate of \snia\
using the parameter \dm15\ (= Phillips parameter) that 
measures the change in $B$-band magnitude during the first 15 days after peak,
the 91bg-like rapid decliners appear as
a distinctive group with \dm15\ $\simgt$ 1.6 mag \citep[e.g.,][]{tauet08, buret14}.
It has been reported that a significant portion, i.e., 15--20~\%, of the entire 
\snia\ population belongs to the group of rapid decliners \citep[e.g.,][]{let11,sriet17}
accompanied by a low peak luminosity in most cases \citep[e.g.,][]{howell01,tauet08}.
While normal \snia\ typically show double-peaked light curves in the $I$ and
and near-infrared bands
where the primary first peak precedes that of the $B$ band,
the rapid decliners show a single $I$-band peak after they reach 
the maximum brightness in the $B$ band \citep[see][and references therein]{tauet08,dhaet17}. 
This behaviour is known to be related to \ni56\ production and
the evolution of temperature and opacity in \snia,
with the 91bg-like group being driven by a smaller amount of \ni56.
These faint and rapidly-declining \snia\ have mostly 
been found in large elliptical galaxies 
and are therefore thought to be associated with old stellar populations
\citep{howell01,sulet06,get11}. 

The distinction of this group of rapid decliners from the normal \snia\ 
in the Phillips relation, however, is less apparent when the decline rate
is represented by the color-stretch parameter (= \sbv) 
defined as the time span between the $B$-band maximum
and that of the $B-V$ color normalized by 30 days \citep{buret14}.
The rapid decliners tend to transition to the post-peak decline phase 
dominated by \co56\ decay in less than 15 days,
which leads to a discontinuity in the slope of the Phillips relation near
\dm15\ $\simeq$ 1.6 mag and makes it
inapplicable for \dm15\ $\simgt$ 1.6 mag.
The disappearance of the clear distinction between Branch Normal and  91bg-like 
based on the \sbv\ parameterization of the decline rate
suggests that the \snia\ in the two groups 
may in fact be similar kinds with a continuous distribution of observables
that appear to be separated by a gap in the number of observed samples between them. 

There is no clear consensus yet on whether or not 
the rapidly-declining \snia, or at least part of them, 
indeed have different origins from the rest.
Some favor different origins for the rapid decliners,
usually relying on the double-degenerate sub-Chandrasekhar (sub-Ch) mass explosions
or the deflagrations in rotating WDs \citep[see][and references therein]{tauet08,dhaet17},
while others have argued that Chandrasekhar-mass explosions can explain at least part of them,
often associated with delayed detonations \citep[e.g.,][]{mrbh07, ashet18}.
Based on the results of extensive radiative transport simulations,
\citet{gk18} reported that 
\snia\ with a large \dm15, i.e., $\simgt$ 1.55 mag, can only be produced by sub-Ch-mass explosions.
This is more or less consistent with the slightly 
broader criterion \dm15\ $\simgt$ 1.4 mag 
obtained by using non-local thermodynamic equilibrium calculations of light curves \citep{bet17}.
In contrast, \citet{het17} reported that a uniform explosion based on Chandrasekhar-mass spherical delayed
detonations can explain most of the observed light curves of \snia\
without the need for a separate mechanism for (extreme) rapid decliners.

A key clue to understanding the origin of the division between
the two groups of Branch Normal and 91bg-like \snia\ lies 
in the nature of the so-called transitional \snia\ straddling them. 
The identification of a statistically significant number of 
such transitional SNe with a continuous distribution of observed properties
could support the common-origin hypothesis.
The light curves of the transitional \snia\ are usually featured with a large Phillips parameter
and a $B$-band peak between two $I$-band peaks.
Their prototypical case is SN~1986G \citep{phiet87} observed 
with \dm15\ $\simeq$ 1.81 mag, 
which is substantially larger than \dm15\ $\sim$ 1.1 mag of Branch Normal,
and a low explosion energy \citep{phiet87,ashet16},
showing a very rapid decay in post-peak evolution.
Spectroscopically, however, it appears to be very similar to normal \snia,
thereby revealing its intermediate nature.
The Phillips parameter of the transitional types lies roughly within the range of
\dm15\ $\simeq$ 1.5--1.8 mag \citep{priet06, ashet16, sriet17, galet18}.
However, transitional types have also been identified in \snia\ beyond this range,
while some \snia\ within the range have been observed with no clear signs 
of intermediate properties \citep[e.g.,][]{tauet08,hsiet15}. 
According to \citet{tauet08}, \snia\ with \dm15\ in the range of 1.75--1.85 mag 
show an inconspicuous secondary $I$-band peak and can be considered as the intermediate 
group between Branch Normal and 91bg-like, 
whereas those in the range of \dm15\ = 1.5--1.75 mag 
featured with a double-peaked $I$-band light curve, 
a relatively small luminosity and a rapid decay
are similar to normal types.
\citet{hsiet15} suggested that \rapd\ \snia\ are 
of transitional nature if they show two long-waveband (i.e., $iYJHK$) peaks in the light curves
with the first and primary peak preceding the $B$-band peak. 
The 10 transitional \snia\ compiled in their work show
\dm15\ $\simeq$ 1.30--1.80 mag and \sbv\  $\simeq$ 0.46--0.86
with average values of 1.62 mag (\dm15) and 0.70 (\sbv).
These transitional \snia\ have been observed mostly 
in non-star-forming galaxies \citep{ashet16, mrbh07},
similar to 91bg-like events (see above).

In this paper, we present multi-color, high-cadence observations of 
the \sni\ \kspn\ (or \iaun) which appears to be a \rapd\ \sni\ 
of transitional nature that is more similar to Branch Normal than 91bg-like,
providing a rare opportunity to investigate
the early photometric evolution of rapid decliners.
In addition, the absence of any apparent host galaxy for this SN
in our deep stack images reaching $\mu_{BVI}$ $\sim$ 28 \mpas\ implies that 
such \snia\ can occur in host galaxy environments that are quite different 
from what have been previously reported.
In \S\ref{sec:obs} below we provide the details of our discovery and monitoring observations of \kspn,
followed by the analyses of the observed light curves and color evolution 
as well as template fitting in \S\ref{sec:lc}.
We compare the observations and model predictions in \S\ref{sec:bol} and \S\ref{sec:pro}:
comparisons of the bolometric light curve with model predictions 
based on different types of \ni56\ distribution (\S\ref{sec:bol}) and
the observed early light curves with model light curves
expected from the interactions between ejecta and companion in \snia\ (\S\ref{sec:pro}).
We show that there is no apparent host galaxy of \kspn\ in \S\ref{sec:host}
and provide the summary and conclusion in \S\ref{sec:sum}. 

\section{Observations}\label{sec:obs}

We conducted high-cadence, multi-color ($BVI$) monitoring observations of 
a 2\degr\ $\times$ 2\degr\ field
containing the galaxy NGC~300 in the direction of the South Galactic pole 
using the Korea Microlensing Telescope Network \citep[KMTNet,][]{ket16} as part of its
KMTNet Supernova Program \citep[KSP,][]{met16,afset19}.
The initial phase of the monitoring started in 2015 July during the commissioning 
test period of the KMTNet and continued until 2017 August 9.
The KMTNet operates three wide-field 1.6-m telescopes located in 
the Cerro-Tololo Inter-American Observatory (CTIO; Chile), 
the South African Astronomical Observatory (SAAO) and the Siding Spring Observatory (SSO; Australia).
All of the three telescopes are equipped with a CCD camera of 
2\degr\ $\times$ 2\degr\ field-of-view (FoV) at 0\farcs4 pixel sampling with Johnson-Cousins $BVRI$ filters \citep{ket16}.
We obtained about 2460 images of the field with 60-s exposure time for 
each of the $BVI$ bands,
and the mean cadence of our observations within the monitoring period
was approximately 3.5 hours for each band.
The typical limiting magnitude for a point source detected with a signal-to-noise (S/N) ratio
greater than 2 in the individual images 
was $\sim$ 22 mag
when seeing was better than 2\arcsec.

\begin{figure}[!t]
\center
\includegraphics{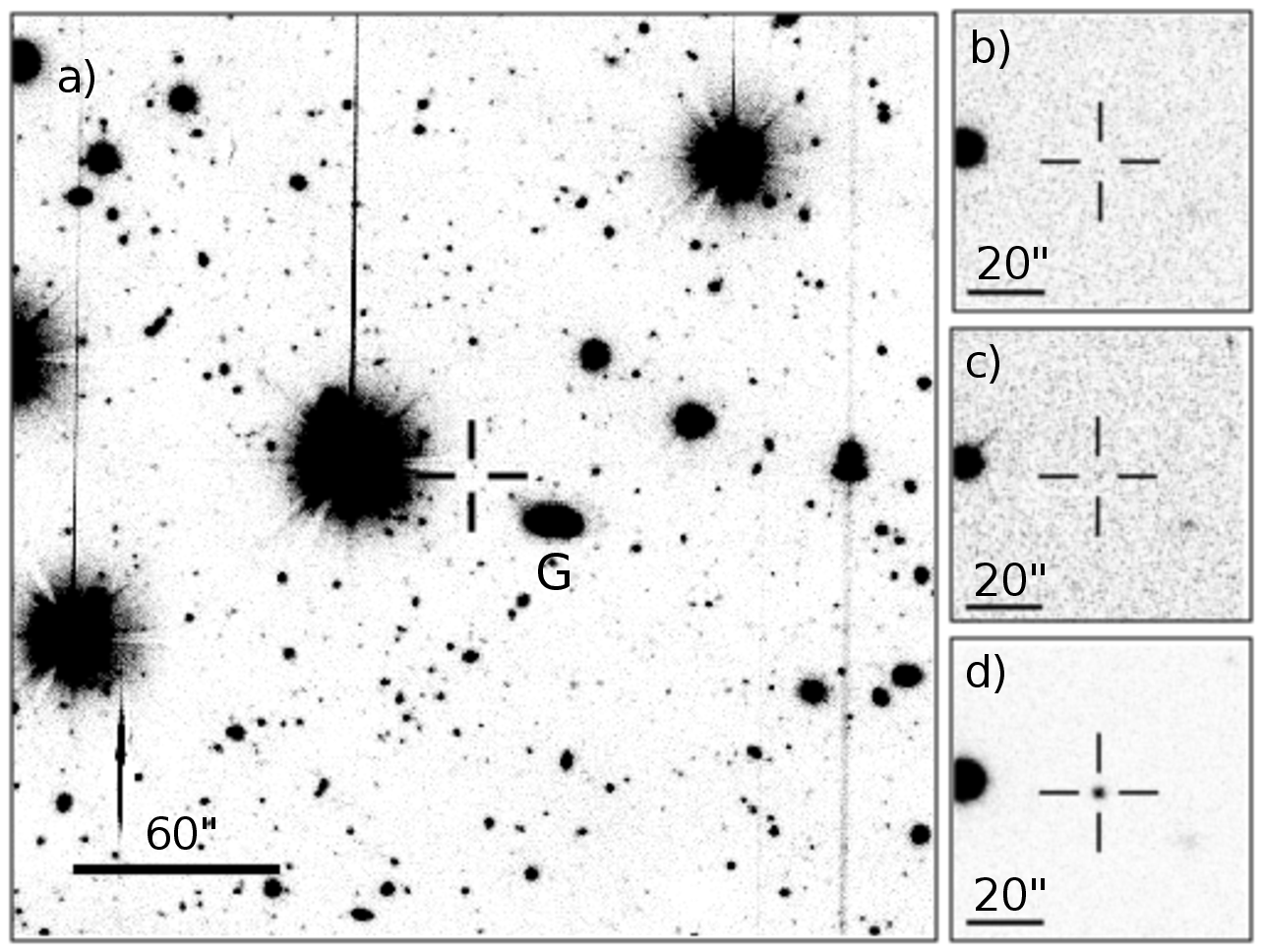}
\caption{(a) Deep KMTNet $I$-band image
centered on the location of \kspn\ obtained by stacking individual exposures taken when the source was below the detection limit.
North is up and east is to the left.
The crosshair marks the location of \kspn, whereas
\galgg\ denotes the apparently extended source 
$\sim$ 27\arcsec\ away in the southwestern direction
from the SN (see \S~\ref{sec:obs} and \ref{sec:host}).
(b)--(d) are individual 60-s $B$-band images focused on the location of \kspn:
(b) the image obtained 7.2 hours prior to the first detection  of the source;
(c) the first detection image obtained at MJD = 57289.89825; 
and (d) the image with the peak observed brightness obtained at 14.1 days from the first detection.
\label{fig:det}}
\end{figure}

A new point source was detected in our $B$-band image obtained 
with the SAAO telescope at 21.56 h on 2015 September 24 (UT), 
or MJD 57289.89825,
with a magnitude of 21.29 $\pm$ 0.37 mag at the coordinate
(RA, decl.) = ($\rm 00^h57^m03.19^s, -37\degr02\arcmin23\farcs6$) (J2000).
The source was subsequently detected in the $V$ band two minutes later with 21.45 $\pm$ 0.33 mag
and in the $I$ band 28 minutes later with 21.59 $\pm$ 0.32 mag, 
while it was not detected in a $B$-band image obtained 7.18 hours before the first detection.
We name this source \kspn.
It remained above the detection limit over the ensuing period of four months at the same location,
reaching observed peak magnitudes of 18.59 $\pm$ 0.02 mag ($B$), 
18.49 $\pm$ 0.02 mag ($V$) and 18.91 $\pm$ 0.05 ($I$) mag over $\sim$ 18 days.
Figure~\ref{fig:det} presents images of the field centered on the location of \kspn.
In the figure, (a) is a deep $I$-band image, 
which reaches a sensitivity limit $\mu_{I}$ $\simeq$ 27.8 \mpas, 
created by stacking 1167 individual images obtained either 
before the first detection of the source or after its disappearance. 
\moon{The image stacking is made using the SWARP package \citep{bet02}.
Each individual frame is subtracted by its own background estimated with the
background mesh size of 512 pixels, and then resampled and median combined
with other frames to create the final stack image.
Only the individual frames whose seeing are better than 2\arcsec\
are used, which can help mitigate the effects of the confusion limit 
in image stacking \citep[e.g.][]{ashcet18}.
There exists no underlying source at the location of \kspn\ in the stack image.}
The three other figures (b)--(d) are the individual 60-s $B$-band images for
the last non-detection exposure obtained 7.18 hours prior to the first detection (b),
the first detection (c), 
and the image from MJD 57304.04163 when the source reached the peak brightness (d). 

The photometric calibration is carried out using more than 10 nearby 
standard reference stars within 15\arcmin\ distance from \kspn\ available in the AAVSO Photometric All-Sky 
Survey\footnote{The AAVSO Photometric All-Sky Survey: Data Release 9, https://www.aavso.org/apass} database. 
Only the reference stars whose apparent magnitudes are in the range of 15--16 mag are used 
to secure high S/N ratios in their flux measurement while avoiding CCD saturation and non-linearity effects.
A local point spread function (PSF), which is obtained by fitting 
a Moffat function \citep{m69, tet01} and sky background emission 
to the reference stars, is fitted for the measurements of fluxes of \kspn\ and the reference stars. 
The AAVSO photometric system consists of the standard Johnson $BV$ and the Sloan $i$ band, i.e., $BVi$.
The calibration of the KMTNet $B$-band data against 
the Johnson $B$-band magnitudes of the AAVSO system
requires a color correction given by
$\Delta B$ $\simeq$ 0.27 ($B-V$) + offset \citep{pet17, pet19}, 
where $\Delta B$ is the $B$-band magnitude differences
between the magnitudes obtained in KMTNet images before color correction
and the standard magnitudes of AAVSO reference stars in the database. 
The same $B$-band color dependence in our data is identified, and
the $B$-band magnitudes of the AAVSO reference stars used in our photometric calibration
are corrected with their known $B-V$ colors.
The final $B$-band magnitudes of \kspn\ are obtained after applying $S$-correction for SNe
by following the procedure described in \citet{stret02}
using template spectra from the \snp\ package (see \S\ref{subsec:temp}). 
No such color dependence has been identified in the KMTNet $V$ and $I$ bands
when calibrated against the AAVSO $V$ and $i$ bands, respectively \citep{pet17, pet19},
which is also confirmed in the data for \kspn.
Thus, the photometric calibration of our KMTNet $VI$-band data are made 
against the AAVSO $Vi$ bands without any color correction. 
As a result, the $BV$-band magnitudes of \kspn\ presented in this paper are
in the Vega magnitude system, while its $I$-band magnitudes are in the AB system.
Table~\ref{tab:mag} contains a sample of the observed magnitudes of \kspn.

According to the dust map of \citet{sfd98}, the total Galactic reddening in
the direction of KSP-OT-201509b is E(\bv) $\simeq$ 0.013 mag. 
Using the updated $R_V$ = 3.1 dust model of \citet{sf11},
we find corresponding extinction of 0.046, 0.034 and 0.021 mag in
the $B$, $V$ and $i$ band, respectively, 
The small extinction is compatible with the location of \kspn\ near the Galactic pole.
Since no host galaxy is identified for the source (see below and \S\ref{sec:host}), 
only this small Galactic extinction is taken into account in our extinction correction.

One notable feature in Figure~\ref{fig:det}(a) is the absence of 
any underlying host galaxy candidate for \kspn\
at its location or in the immediate vicinity of the source in our deep $I$-band image.
The absence is also confirmed in the $B$ and $V$ bands.
The most conspicuously extended nearby object of \kspn\ is
located at (RA, decl.) $\simeq$ ($\rm 00^h57^m01.23^s, -37\degr02\arcmin37\farcs5$) (J2000),
$\sim$ 27\arcsec\ away in the southwest. 
We call this object \galgg\ hereafter as we denote in the figure.
In order to measure the redshift and understand the nature of \galgg, we conducted 
spectroscopic observations of the source using
the Low Dispersion Survey Spectrograph 3 (LDSS3)\footnote{http://www.lco.cl/telescopes-information/magellan/instruments/ldss-3}
on the 6.5-m Magellan Clay telescope on 2016 August 11.
We used the VHP-All grism coupled with 1\arcsec\ slit for dispersion
and obtained a single 600~s exposure.
See \S\ref{sec:host} for the spectrum of \galgg\ and its analysis.

\medskip
\section{Light Curves and Classification of \kspn\ as a Rapidly-Declining Transitional \sni}\label{sec:lc}

\subsection{Light Curves and Epoch of the First Light}\label{subsec:lc}

Figure \ref{fig:lc} shows $BVI$-band light curves of \kspn\ 
obtained with the KMTNet for about two months.
Over approximately the first two week, 
the source gradually ascends to its peak brightness more than three 
magnitudes brighter than that of the first detection.
The observed peak brightnesses are 18.59, 18.49 and 18.91 mag for
$B$, $V$, and $I$ band, respectively.
(See Table~\ref{tab:par} for the observed and estimated parameters of \kspn.)
There exist clear band-dependent differences 
in the manner of how post-peak decay evolves in Figure~\ref{fig:lc}.
The $B$-band decay is much faster than those of the other bands, 
while the $I$-band light curve is featured with a secondary peak about two 
weeks after the first and primary peak, which precedes the $B$-band peak.
We estimate the average strength of the secondary $I$-band peak 
within 20--40 days after the epoch of the $B$-band peak 
to be $<I>_{\rm 20-40}$ = 0.309 $\pm$ 0.002
with respect to the maximum brightness of the first and primary
peak of the $I$ band \citep[][]{kriet01}. 
The overall evolution of the observed light curves of \kspn\ -- 
which includes 
the initial ascent in brightness,
the presence of a secondary peak in the $I$ band,
the absence of any apparent post-peak plateau or linear phase, and
the slow decay -- 
identifies \kspn\ as a \sni\ powered by radioactive decay 
of \ni56\ and \co56\ \citep[e.g.,][]{het96}. 
We compare in Figure~\ref{fig:comp} the $I$-band light curve (black) of \kspn\
with those of four well-sampled \snia: 
SN2011fe (red), SN1994D (blue), SN2005ke (cyan) and SN2005bl (green).
The first two are Branch Normal with \dm15\ $\simeq$ 1.1 and 1.4 mag, respectively, whereas 
the last two are 91bg-like with more rapid post-peak decline rates 
\dm15\ $\simeq$ 1.7 (SN2005ke) and 1.9 (SN2005bl) mag.
The post-peak decline rate of \kspn\ in Figure~\ref{fig:comp}
appears to be intermediate between the two groups. 

Figure~\ref{fig:epo} presents the observed early light curves of \kspn\ 
prior to 8 days before the peak, normalized by its peak brightness.
Power laws have been adopted to describe the evolution of early light curves of \snia\
when intensities are substantially 
lower \citep[e.g., $<$ 40~\%, see][]{ollet15} than
the peak intensity since the brightness of a \ni56-powered expanding homologous sphere
mediated by photon diffussion process is expected to be $\propto$ $t^2$ where $t$ is time
\citep[e.g.,][]{nuget11}.
We fit the early light curves of \kspn\ during the first $\sim$ 10 days 
of its detection in Figure~\ref{fig:epo} 
with a power law $F(t)$ $\propto$ $(t-t_0)^{\alpha}$,
where $F$ and $t_0$ represent the observed brightness 
and the epoch of the first light, respectively.
The best-fit power-law indices are 
$\alpha$ = 2.0 $\pm$ 0.2 ($B$), 1.9 $\pm$ 0.2 ($V$) 
and 2.1 $\pm$ 0.2 ($I$) 
and the epoch of first light is \t0\ =  --18.4 $\pm$ 0.6 days 
since the peak in the observer frame, 
or --17.2 $\pm$ 0.5 days in the rest frame.
(See \S\ref{subsec:temp} for the estimation of the epoch of the peak brightness 
and the redshift of the source.)
The fitted power-law indices are $\simeq$ 2 as expected for a homologous expansion, 
similar to what have been found
in other \snia\ \citep[e.g.,][]{nuget11,ollet15,dimitriadis19}.
\moon{ We also estimate the epoch of the first light using
a Gaussian process extrapolation which can provide 
a less model-dependent inspection of the epoch.
The Gaussian process extrapolation-based epoch of the first light 
is $-$16.4 $\pm$ 0.5 days (rest frame) for the $BVI$ light curves,
which is slightly smaller but consistent with $-$17.2 $\pm$ 0.5 days  
from our power-law fitting. We adopt $-$17.2 $\pm$ 0.5 days as the
epoch of the explosion of \kspn\ in its rest frame.}

\subsection{Color Evolution}\label{subsec:color}

Figure~\ref{fig:color} shows the evolution of the $B-V$ and $V-I$ colors of
\kspn\ aligned with that of the $I$-band light curve.
The three vertical dotted lines mark three color epochs over which
the colors show a notable phase transition in their evolution.
The first and third color epochs are --4.5 and +18 days, respectively, from
the epoch of the $B$-band peak brightness estimated using SN template fitting of the 
observed $BVI$-band light curves (see below for the details).
These two color epochs roughly correspond to those of the primary and secondary $I$-band peaks, respectively. 
The second color epoch is +7 days which is close to the midpoint
between the first and third epochs,
near the onset of the $I$-band secondary rise.

We summarize the evolutions of the \bv\ and \vi\ colors,
which are largely synchronous with that of the $I$ band, as follow.
\begin{itemize}
\itemsep0em 
\item  Prior to the first color epoch of --4.5 days,
the \bv\ color starts at $\simeq$ 0.2 mag around --14 days,
or $\sim$ 3.3 days in the rest frame since the epoch of first light (Table~\ref{tab:par}), 
and slowly becomes bluer to $\simeq$ 0.0 mag in about 10 days
near the first color epoch. 
The blueward evolution of the \bv\ during this early phase of \kspn\ 
appears to be consistent with that of the early red population of \snia,
which comprises predominantly Branch Normal types \citep{stret18}.
It is thought to be due mainly to the heating  from \ni56\ radioactive decay as 
the deeper layer of the ejecta is revealed \citep[see][]{het17}.
Indeed, we show in \S\ref{sec:bol} that the observed early light curves of \kspn\ 
can be explained by a shallow \ni56\ distribution  \citep[e.g.,][]{pm16}
leading to an early increase in color temperature.
The \vi\ color of \kspn\ during this phase, 
on the other hand, shows a redward evolution
from approximately --0.5 to --0.1 mag within the first two days 
before it stalls relatively flat until the first color epoch.
SNe in general have rarely been observed with \vi\ colors at these early epochs, 
and the origin of the early redward evolution of \kspn\ is unclear. 
It is conceivable, however, that the evolution of spectral features 
-- e.g., Ca II features in the $I$ band \citep[][see Figure~1 therein]{paret12} --
are responsible for it.
\item Across the first color epoch of --4.5 days, 
when the $I$-band light curve nears its primary peak, 
both colors change the direction of their evolution 
in that \bv\ becomes redder while \vi\ becomes bluer.
We attribute this redward evolution in \bv\ to the absorption 
by iron peak elements \citep[e.g.][]{het17},
whereas the blueward evolution in \vi\ 
to temperature increase.
\item The \bv\ color evolves redward relatively monotonically between
the first (i.e., --4.5 days) and third (i.e., +18 days) color epochs,
which is largely equivalent to the period between the two $I$-band peaks,
by $\simeq$ 1.2 mag,
whereas the \vi\ color changes its direction of evolution again around the 
second color epoch (i.e., +7 days) from blueward to redward until
it reaches the third color epoch. 
This is due to the recombination process of Fe III leading to 
a rebrightening in the $I$ band by redistribution of blue/UV radiation
\citep[e.g.,][see also \S\ref{sec:intro}]{kas06}.
\moon{We note that the Fe III recombination also plays a critical role
in determining the color evolution of \snia\ in the near-infrared wavebands
with respect to their light curves \citep{dhaet15}, similar to what
we report here.}

\item After the third color epoch, both colors 
gradually evolve blueward as the SN enters the Lira law phase known
to have a linear blueward evolution for an extended period \citep{phiet99}.
\item The \bv\ and \vi\ colors at the peak epoch
in Figure~\ref{fig:color} are 0.056 and --0.34 mag, respectively.
According to \citet{tauet08},
the observed \bv\ color of \kspn\ at the peak epoch 
is significantly bluer than 
the 0.4--0.7 mag range known for 91bg-like,
while it is consistent with the value
expected for a \sni\ with \dm15\ $\simlt$ 1.75 mag.
\item \snia\ with \dm15\ in the range of 1.5--1.75 mag
are known to show an initial blueward evolution between --10 and 10 days 
in \vi\ followed by a rapid redward evolution \citep{tauet08} as observed in \kspn,
while this initial blueward evolution is absent in 91bg-like.
It is quite noteworthy, however, that the \vi\ color evolution of \kspn\ 
in Figure~\ref{fig:color} also reveals that the initial blueward evolution 
(between --10 and 10 days) is in fact preceded by an earlier 
redward evolution at $<$ --10 days as explained above.
A similar \vi\ color pattern  at comparably early epochs has been observed 
in SN~1994D which is a normal but relatively over-luminous  \sni\ \citep{patet96}.
\item Overall, the observed color evolutions of \kspn\ are 
compatible with what can be expected from a \sni\ 
with \dm15\ = 1.62 mag as measured for the source (see \S\ref{subsec:temp})
and appear to be more similar to Branch Normal than 91bg-like.
Also, the similarity of its color evolutions to those observed
in other \snia\ indicates that there is an insignificant extinction 
by its potential host galaxy as we assumed (\S\ref{sec:obs}).
\end{itemize}

\subsection{Template Fitting and Parameters of \kspn}\label{subsec:temp}

\begin{figure}[ht!]
\includegraphics[width=\textwidth]{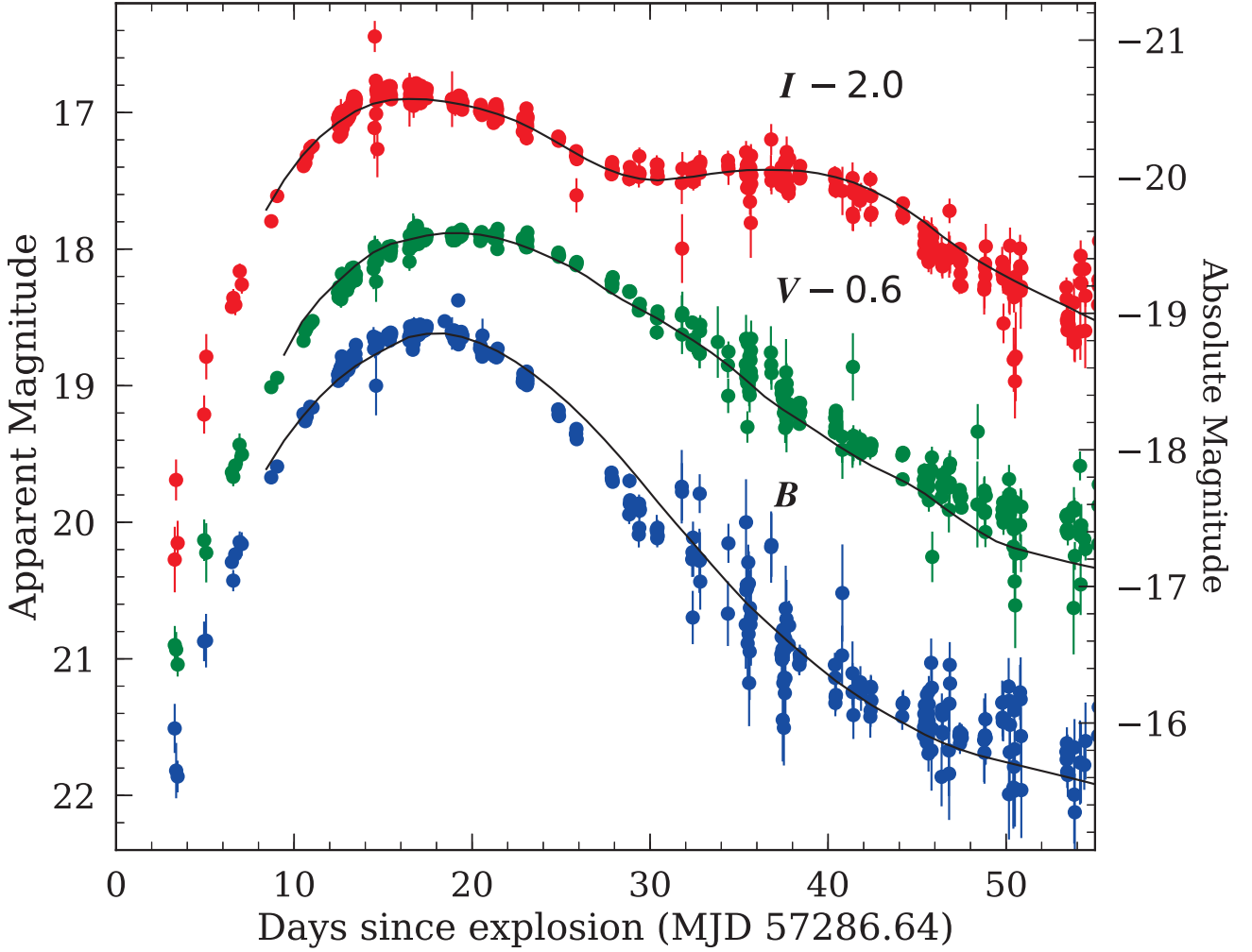}
\caption{$BVI$ light curves of \kspn\ observed with the KMTNet: 
blue, green and red circles for the 
$B$, $V$, and $I$ band, respectively. 
The abscissa represents days from the estimated explosion epoch
(MJD 57286.64) of the source,
which is adopted to be equal to the epoch of the first light
(see \S\ref{subsec:lc} and Table~\ref{tab:par}),
in the observer frame;
the ordinate does the observed apparent magnitudes (left) and 
corresponding absolute magnitudes (right) 
at the distance modulus of 37.47 mag (see \S\ref{subsec:temp}).
The $V$- and $I$-band magnitudes are shifted vertically
by --0.6 and --2.0 mag, respectively,
to separate the overlapping light curves of the three bands. 
For the several initial epochs at the beginning, 
adjacent individual exposures are binned together to increase S/N ratios.
The black solid line represents the best-fit 
template to the entire $BVI$-band light curves obtained 
from \snp\ fitting after $S$-correction with $z$ = 0.072 (see \S\ref{subsec:temp}).
\label{fig:lc}}
\end{figure}

\begin{figure}[ht!]
\includegraphics{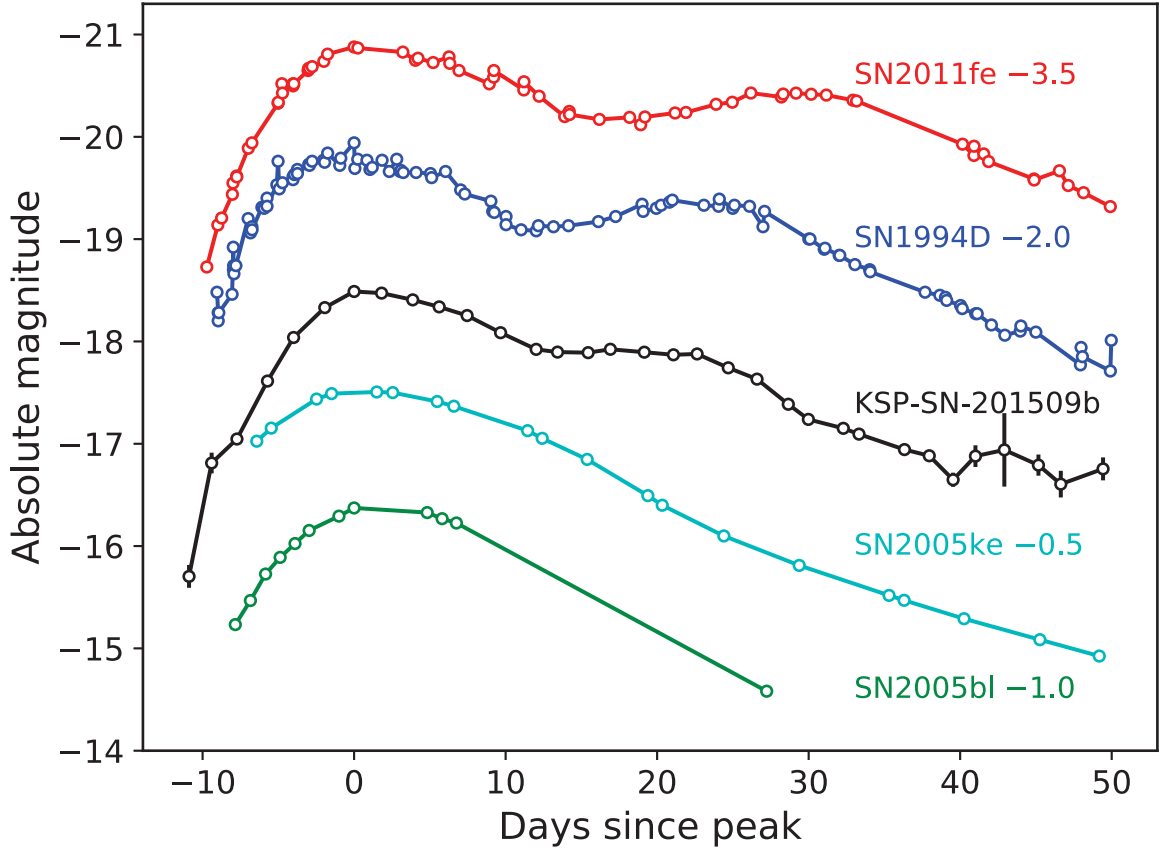}
\caption{Comparison of the $I$-band light curve of \kspn\ (black, middle) 
with those of four other well-sampled \snia\
in the increasing order of the post-peak decline rate from top to bottom.
The light curves are aligned with the epochs of their respective  
$I$-band peak in the source rest frame.
The top two light curves are for 
SN2011fe (red, shifted by --3.5 mag) and 
SN1994D (blue, shifted by --2.0 mag) \citep[][and references therein]{pet13,wyz95}.
They have \dm15\ $\simeq$ 1.1 and 1.4 mag, respectively.
The bottom two light curves are for  
SN2005ke (cyan, shifted by --0.5 mag) and
SN2005bl (green, shifted by --1.0 mag) 
that have \dm15\ $\simeq$ 1.7 and 1.9 mag, respectively
\citep{conet10}.
\label{fig:comp}}
\end{figure}

\begin{figure}[ht!]
\includegraphics{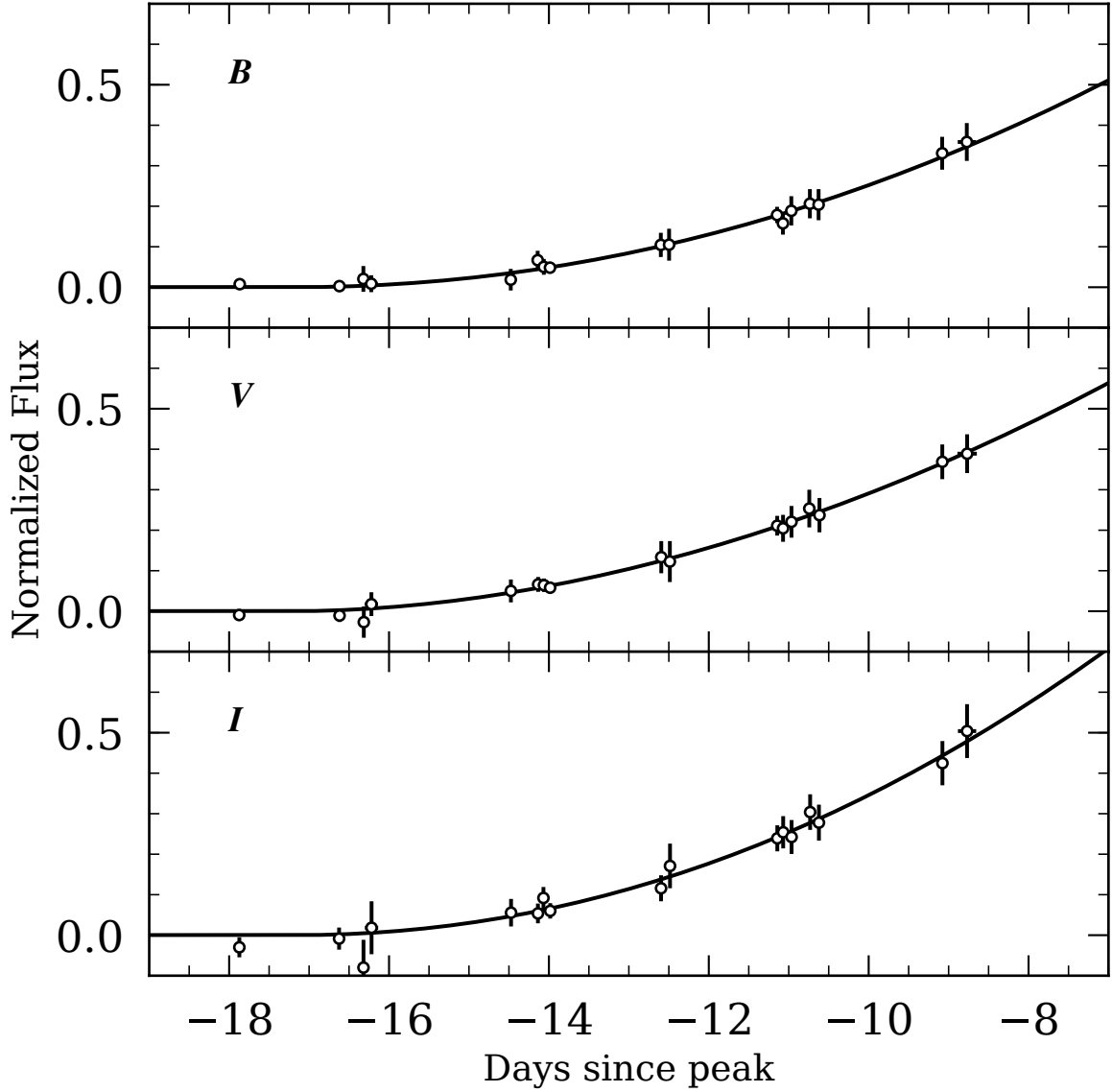}
\caption{Early $BVI$ (from top to bottom) light curves 
(open circles) of \kspn\ normalized by the peak intensity of each band.
Three to seven adjacent individual flux measurements, 
each obtained with 60-s exposure time,
are binned together to increase S/N ratios, and 
the error bars correspond to 95~\% confidence level 
of the flux measurement. 
The solid lines are the 
best-fit power laws obtained for each band.
The abscissa represents days from the epoch of the peak $B$-band
brightness estimated by \snp\ fitting in the source rest frame.
}
\label{fig:epo}
\end{figure}

\begin{figure}[ht!]
\includegraphics{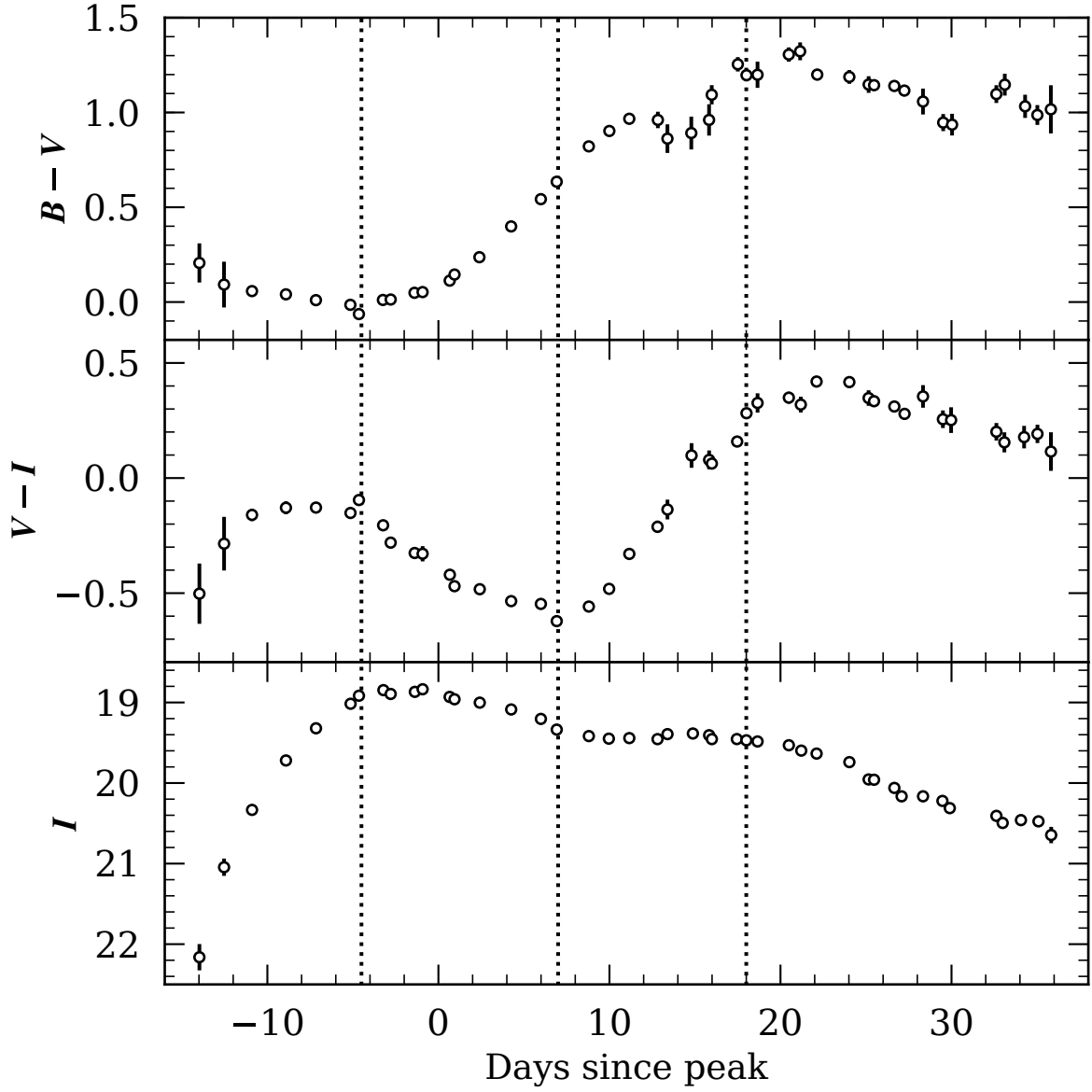}
\caption{Evolution of the \bv\ (top panel) 
and \vi\ (middle panel) colors of \kspn\ aligned with 
the $I$-band light curve (bottom panel).
The data are binned up to 1 day interval to increase S/N ratio 
and the error bars represent the uncertainties 
measured at 68~\% confidence level.
The abscissa represents days from the epoch of the peak $B$-band
brightness estimated in \snp\ fitting in the source rest frame.
The three vertical lines mark three notable color epochs of --4.5, 7 and 18 days. 
\label{fig:color}}
\end{figure}

In order to estimate key physical parameters of \kspn\ as a \sni, 
we conduct template fitting of its observed light curves  (Figure~\ref{fig:lc})  
using two \sni\ light curve templates from the SN template fitting package \snp\ \citep[v181212;][]{buret11},
one for Branch Normal and the other for 91bg-like.
Because no spectroscopic observations were made for \kspn\ 
(due to its detection during the commissioning test period of the KMTNet)
and its host galaxy is unknown (see \S\ref{sec:host}),
we estimate its distance using an iterative template model fitting technique.
For this, we adopt a set of 100 trial redshifts in the range  $z$ = 0.03--0.1
for potential redshifts of the source.
This corresponds to the distance modulus (DM) of $\simeq$ 35.6--38.6 mag
based on the cosmological model of \citet{rieet16} with parameters 
$H_0$ $\simeq$ 73.24 \kms\ Mpc$^{-1}$,  
$\Omega_{\rm M}$ $\simeq$ 0.27 and $\Omega_{\rm \Lambda}$ $\simeq$ 0.73.
\moon{Note that these cosmological parameters are more recently updated values 
that are slightly different from those adopted in the \snp\ package internally, 
which are  $H_0$ $\simeq$ 72 \kms\ Mpc$^{-1}$, $\Omega_{\rm M}$ $\simeq$ 0.28 
and $\Omega_{\rm \Lambda}$ $\simeq$ 0.73 \citep{speet07}.} 
Given $B$ = 18.59 $\pm$ 0.02 peak apparent magnitude of \kspn\ (Table~\ref{tab:par}),
this redshift range ensures us to investigate the full range of 
peak absolute magnitudes typically displayed by \snia,
which is between --17 mag and --20 mag as shown in Figure~\ref{fig:phil},
for conceivable peculiar motion of the source (see below).
We use \snp\ 
to conduct SN template fitting and $K$-corrections with the input redshifts, assuming no host galaxy extinction.

Figures~\ref{fig:LCtemplate} and \ref{fig:LCzfitVi} show the results
of our template fitting to the light curves of \kspn. 
In Figure~\ref{fig:LCtemplate} we compare the observed $V$-band (top panel)
and $I$-band (bottom panel) light curves of the source
with the best-fit template light curves for Branch Normal (solid line) and 91bg-like (dashed line).
We exclude the $B$-band light curve in this initial template fitting
stage to avoid uncertainties involved in the 
$S$-correction process (see \S\ref{sec:obs}).
While the best-fit $V$-band light curve from the Branch Normal template 
in the figure appears to give a good match to the observed light curve, 
that from the 91bg-like template slightly overpredicts the decline rate
immediately after the peak. 
This discrepancy is more obvious in the $I$ band where 
the double-peaked light curve of \kspn\ is apparently 
incompatible with the best-fit 91bg-like template,
showing that the source is more similar to Branch Normal than 91bg-like.

Figure~\ref{fig:LCzfitVi} presents the distribution of the fitted 
parameters from the template
fitting as a function of the trial redshifts (bottom abscissa),
or corresponding cosmological DM (top abscissa).
The top panel shows $\Delta$(DM) (left ordinate), 
which is the offset between the \snp-fitted DM at a given redshift and the cosmological DM from the top abscissa, 
representing the peculiar motion (as calculated in the right ordinate) of \kspn\
if the fitted DM and assumed redshift are indeed the real values for the source.
As in the figure, no peculiar motion of \kspn\ is required if the source is 
located at $z$ = 0.072 for Branch Normal (solid line) or 0.077 for 91bg-like (dashed line).
\moon{At redshifts smaller than 0.06 or larger than 0.08, 
in order for the input redshift and the cosmological redshift 
from the distance modulus obtained in the SNooPy fitting to be comparable,
the required peculiar motions
become significantly larger, i.e., $>$ 1000 \kms, than what can be expected from \kspn\ (see below)
for the Branch Normal template fitting. This also applies to the 91bg-like template fitting,
but at slightly larger redshift values.}

In the bottom panel of Figure~\ref{fig:LCzfitVi},
the distribution of the best-fit \csp\ \sbv\ (left ordinate) 
increases along the trial redshifts as the quality of the fitting becomes poorer.
The growing incompatibility between the template and observed light curves  at a higher redshift,
caused mainly by the secondary $I$-band peak,
results in larger best-fit values of \csp\ for higher trial redshifts. 
The Phillips parameters (\dm15, right ordinate) in the panel 
are obtained using the relation between \sbv\ and \dm15\ in \citet{buret14}.
At $z$ = 0.072, the best-fit \sbv\ is 0.75 (or \dm15\ = 1.54 mag) for Branch Normal,
while it is 0.82 (or \dm15\ = 1.39 mag) for 91bg-like at $z$ = 0.077.
The \chisqr\ values of the best-fit templates at these redshifts are
$\simeq$ 2 for Branch Normal ($z$ = 0.072) 
and $\simeq$ 3 for 91bg-like ($z$ = 0.077).
Overall, the best-fit Branch Normal templates give 
systematically, although slightly,
smaller \chisqr\ values of \simlt\ 3
for all the trial redshifts.
At higher redshifts $z$ \simgt\ 0.08, however, 
\chisqr\ values of the best-fit 91bg-like templates
become increasingly greater than 3.

Based on the results shown in Figures~\ref{fig:LCtemplate} 
and \ref{fig:LCzfitVi} from the \snp\ template fitting,
we estimate the redshift of \kspn\ to be $z$ = 0.072 $\pm$ 0.003 and 
conclude that the source is closer to Branch Normal than 91bg-like  as follows. 
First of all, the comparison between the observed and best-fit templates (Figure~\ref{fig:LCtemplate}),
especially the $I$-band comparison (bottom panel),
gives clear preference for Branch Normal for \kspn.
Also, the best-fit color stretch parameter \sbv\ $\simeq$ 0.82 for 
the 91bg-like template is unacceptably larger than those found in 91bg-like SNe,
while the best-fit \sbv\ $\simeq$ 0.75 for Branch Normal is largely acceptable, 
although somewhat small, for a Branch Normal SN.
This shows that the 91bg-like template is apparently incompatible with 
the observed light curves of \kspn\ at an admissible redshift range. 
The estimated redshift of \kspn\ is largely derived from the conceivable range 
of reasonable proper motions for its progenitor. 
As explained in \S\ref{sec:host}, 
it is highly likely that the host of \kspn\ is an unidentified 
dwarf galaxy whose brightness is fainter than $\sim$ 25 mag in $BVI$.
In the local environment around the Milky Way, the peculiar velocities of dwarf galaxies are at the level of 200 \kms\ \citep{let07}, 
with a few occasional outliers showing higher velocities
in special environments \citep[e.g.,][]{dgc01}. 
For a galaxy like the Milky Way, $\sim$ 500 \kms\ is the escape velocity,
while rare objects with faster velocities of up to $\sim$ 1000 \kms\ are considered 
hypervelocity stars \citep{brown15, beei17}.
\moon{Adopting 500 \kms\ as a conservative, or upper-limit, assessment of 
the potential peculiar motion of \kspn,
we obtain $\Delta z$ = 0.003 as the uncertainty of the best-fit redshift 
from the Branch Normal template fitting 
after taking the uncertainty in the DM measurement
from the \snp\ fitting, which includes
the uncertainty contributions  from the \snp\ template calibration
as well as from adopting the more recently updated cosmological
parameters than \snp\ (see above),
and the uncertainty corresponding to 500 \kms\ peculiar motion into account together.
The $\sim$ 4\% level of the photometric redshift uncertainty is 
comparable with what have been found in some other 
photometric studies of \snia\ \citep[e.g.,][]{palet10}.}
This gives $z$ = 0.072 $\pm$ 0.003 and DM = 37.47 $\pm$ 0.10 mag, 
or luminosity distance of 311.9 Mpc,
for the source (Figure~\ref{fig:LCzfitVi}).
The small difference between the best-fit redshifts of Branch Normal and 91bg-like
indicates that any potential systematic uncertainties resulting from applying the
Branch Normal template to KSP-OT-201509b, which shows transitional nature (\S\ref{subsec:tran}), 
for its redshift determination is small.

We now conduct $S$-correction of the $B$-band light curve 
using this redshift and the spectral time series of the best-fit 
Branch Normal template shown in Figure~\ref{fig:LCtemplate}
as identified in the $VI$-band light curve analyses above.
({\it Note that the $B$-band light curve in Figure~\ref{fig:lc} is the one 
obtained after $S$-correction.})
We then use SNooPy with the entire $S$-corrected $BVI$ light curves 
to obtain the best-fit Branch Normal template for \kspn\
(Figure~\ref{fig:lc}, black lines) 
from which we derive $K$-corrections. 
We adopt the epoch of the $B$-band peak brightness \tp\ = 57305.08 $\pm$ 0.03 (MJD) of the best-fit template,
which is 15.18 $\pm$ 0.03 days after the epoch of the first detection
and $\sim$18.4 days after the epoch of first light, 
as the peak epoch of \kspn\ from the template.
Using the $S$- and $K$-corrected $BVI$ curves of the SN in the rest frame, 
we estimate absolute peak magnitudes  of 
--18.94 $\pm$ 0.01 ($B$), --18.93 $\pm$ 0.01($V$) and --18.38 $\pm$ 0.01 ($I$) mag 
with their respective epochs. 
We finally estimate the \csp\ and \pp\ of \kspn\ to be 
\sbv\ = 0.54 $\pm$ 0.05 using the 
$B$-band light curve and 
$B-V$ color evolution of the source as defined in \citet[][]{buret14} and
\dm15\ = 1.62 $\pm$ 0.03 mag by conducting polynomial fitting 
of the $B$-band light curve in the rest frame of $z$ = 0.072.
These final \sbv\ and \dm15\ values are different from those obtained 
from the $VI$-band template fitting above
(i.e., \sbv\ $\simeq$ 0.75 and \dm15\ $\simeq$ 1.54 mag for Branch Normal;
\sbv\ $\simeq$ 0.82 and \dm15\ $\simeq$ 1.39 mag for 91bg-like),
and we attribute the discrepancy to the transitional nature of \kspn\ 
(see below)
since neither template is an intrinsically good match to the observed light curves of the source.
See Table~\ref{tab:par} for the compilation of the physical parameters of \kspn.

\moon{As described above, the redshift of \kspn\ is estimated using its observed peak brightness
by allowing reasonable peculiar motion of the source in the frame of the \snp\ template fitting, rather than purely based on the goodness of the light-curve fitting.
The minimum \chisqr\ values of the \snp\ template fittings change only slightly, 
i.e., mostly between 2 and 3,
over the trial redshift range of 0.01--0.1, making the parameter 
inadequate to discern the redshift of the source alone.
In order to examine the validity of our redshift estimation from the \snp-based analyses,
we apply two other \snia\ light curve fitting packages to the observed light curves of \kspn.
We first use the SALT2 fitting package \citep{guyet07} with the same trial redshift range
of 0.01--0.1, which results in only a slight variation, i.e., between 4 and 5,  of the minimum \chisqr\ values,
as we find in the \snp\ fitting.
Both the best-fit redshifts from the SALT2 fitting of the $B$- and $V$-band 
light curves determined by the minimum \chisqr\ value are consistently $\sim$ 0.07, 
similar to what we obtain for \kspn\ from the \snp-based analyses above, 
while it is $\sim$ 0.02 for the $I$ band.
This significant discrepancy between the $BV$ and $I$ band might be due to 
incompatibility between the SALT2 package and the transitional nature of \kspn,
which is expected to affect the $I$-band fitting more than the other bands,
although it is important to note that these best-fit redshifts are 
poorly constrained given the small variations in the minimum \chisqr\ values.
Figure~\ref{fig:SLAT2}, which shows the distribution of the best-fit shape parameter (= \x1)
from our SALT2 fitting of the \kspn\ light curves over the trial redshift range of 0.01--0.1,
provides a consistency check between the \snp- and SALT2-based fitting results.
In the figure, the two shaded areas represent the ranges of the redshift and \x1\ 
expected from the \snp\ fitting for \kspn: the vertically shaded area is $z$ = 0.072 $\pm$ 0.003
that we estimate as the redshift of the source,
while the horizontally shaded area is \x1\ = $-$1.89 $\pm$ 0.05 converted from
the fitted \sbv\ values in Figure~\ref{fig:LCzfitVi} (bottom panel) for the redshift
using the known relationship between \x1\ and \sbv\ in \citet{buret14}.
The SALT2-based \x1\ parameter intersects with the two shaded areas, 
showing that the results of the \snp\ fitting are consistent with 
those of the SALT2 fitting on the known relation between \sbv\ and \x1.
In addition to the SALT2 fitting, we also apply the SiFTO SN fitting package \citep{conet08} 
to the observed light curves of \kspn\ over the trial redshift range of 0.01--0.1. 
Although the best-fit redshift from the SiFTO fitting is poorly constrained again 
due to the small variation in the minimum \chisqr\ values, we obtain $\sim$ 0.07 
for the best-fit redshift, compatible with \snp\ and SALT2 fitting results.}

\subsection{Transitional Nature of \kspn}\label{subsec:tran}

\begin{figure}[ht!]
\includegraphics{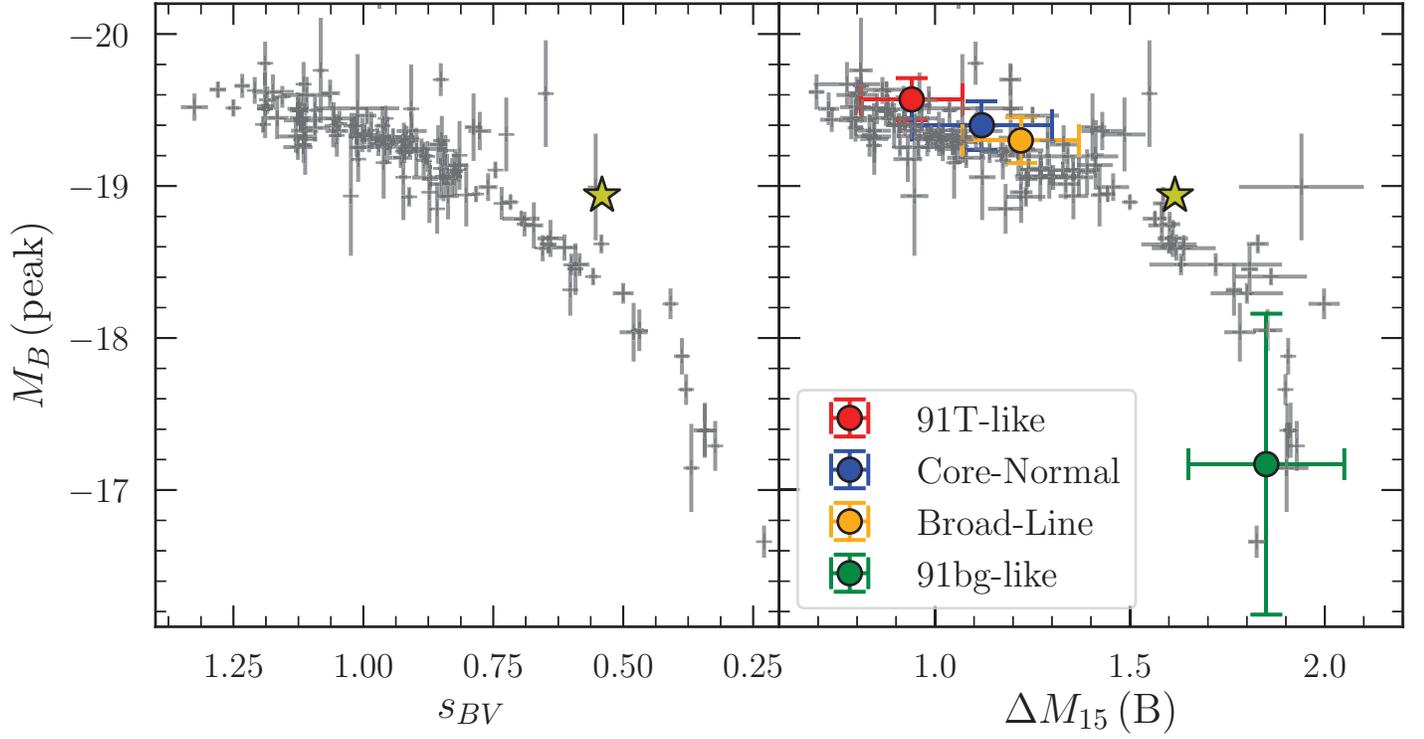}
\caption{({\it Left Panel})
Distribution of \sbv\ and the $B$-band peak absolute magnitude (crosses)
of \snia\ compiled in \citet{buret18} with those of \kspn\ (yellow star).
({\it Right panel}) Same as the left panel, but for \dm15.
Also shown are the mean values of the four major subtypes of \snia\ from \citet{pfp14}:
91T-like (filled red circle),
Core-Normal (filled blue circle), Broad-Line (filled orange circle), 
and 91bg-like (filled green circle).}
\label{fig:phil}
\end{figure}

\begin{figure}[ht!]
\includegraphics{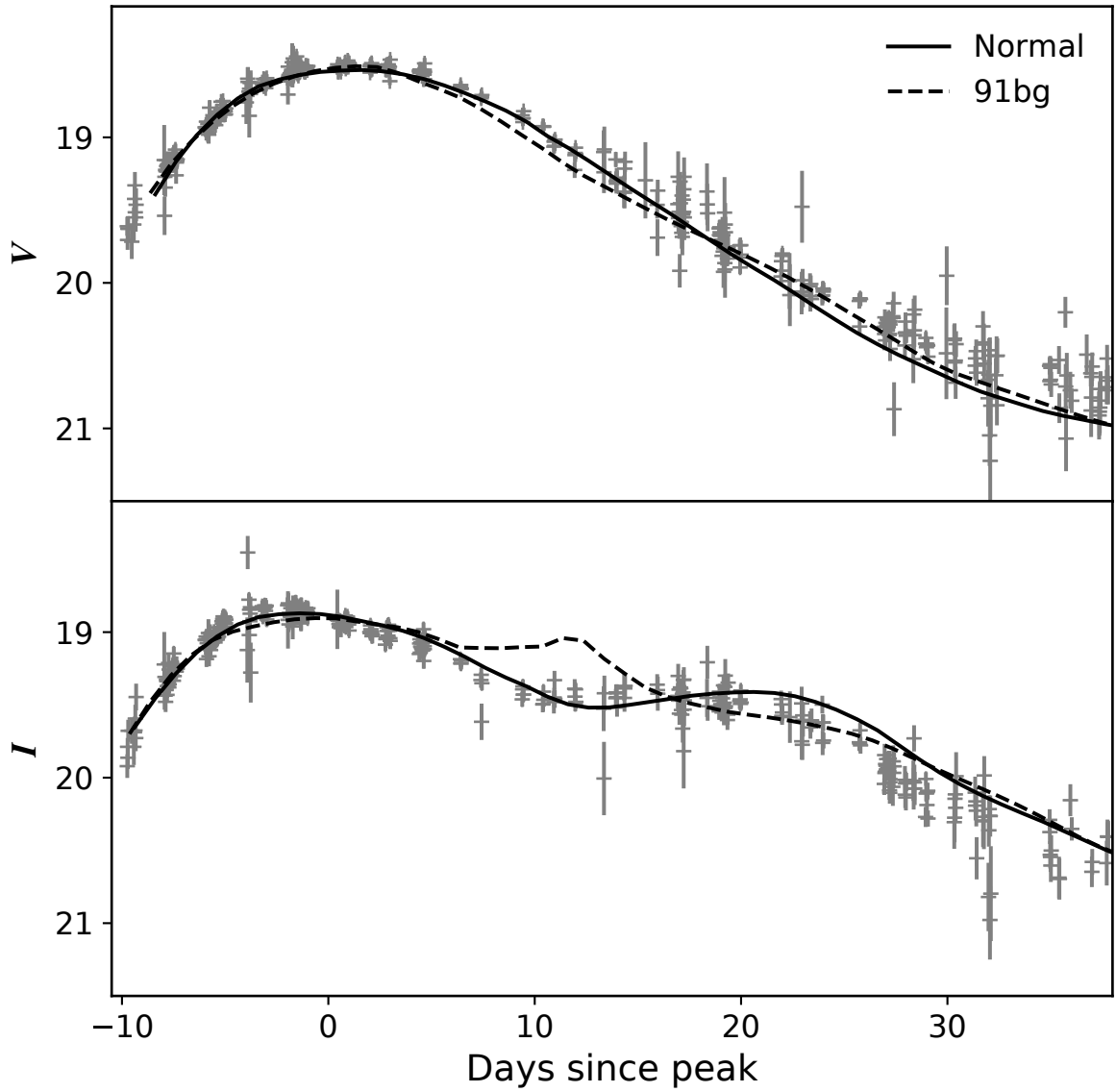}
\caption{({\it Top Panel}) Observed $V$-band light curve of \kspn\
overlaid on that of the best-fit \snia\ Branch Normal template (solid line, $z$ = 0.072) and 
91bg-like template (dashed line, $z$ = 0.077) obtained from \snp\ fitting.
The abscissa represents days from the epoch of the peak $B$-band
brightness estimated in \snp\ fitting in the observer frame.
({\it Bottom Panel}) Same as the top panel, but for the $I$-band light curve of \kspn.}
\label{fig:LCtemplate}
\end{figure}

\begin{figure}[ht!]
\includegraphics[width=\textwidth]{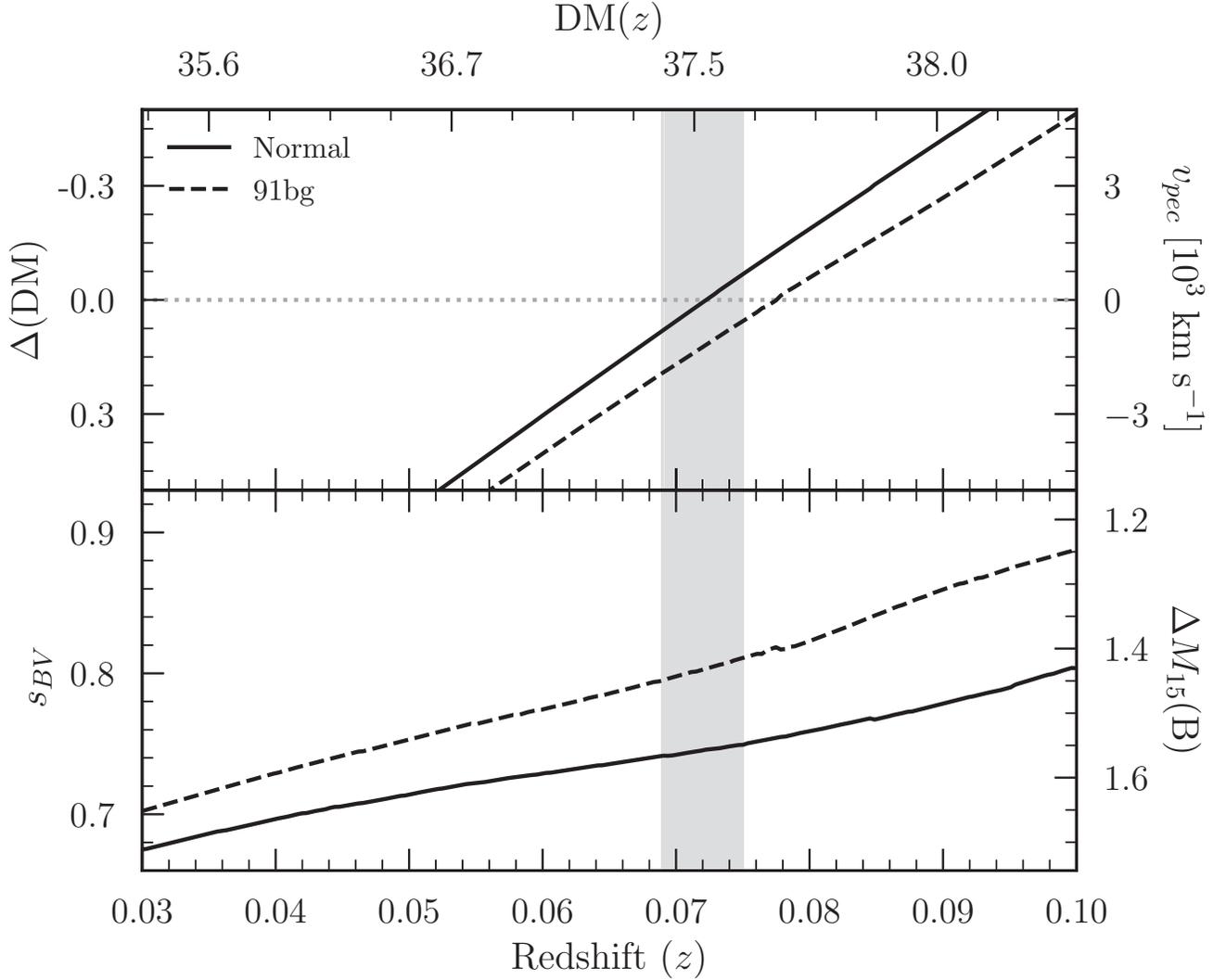}
\caption{Distribution of template fitting parameters
of the \kspn\ light curves in \snp\
as a function of 100 trial redshifts in the range $z$ = 0.03--0.10.
The abscissae represent the trial redshift (bottom)
and the corresponding cosmological distance modulus 
(top), DM($z$), 
consistent with the cosmological model of \citet{rieet16}.
The shaded region highlights the inferred redshift range $z$ = 0.072 $\pm$ 0.003 of the source.
({\it Top Panel}) Distribution of the offset, $\Delta$(DM) (left axis), 
between the best-fit distance modulus in \snp\ and the cosmological distance modulus for a given redshift
and the required peculiar velocity, $\varv_{pec}$ (right axis), reconciling the distance modulus offset are shown.
The solid line is for Branch Normal template fitting; 
the dashed line is for 91bg-like template.
({\it Bottom Panel}) Same as the top panel, but for the best-fit \sbv\ (left axis)
at a given redshift and \dm15\ (right axis) calculated with \sbv\ using the relation 
between the two parameters in \citet{buret14}.}
\label{fig:LCzfitVi}
\end{figure}

\begin{figure}[ht!]
\includegraphics{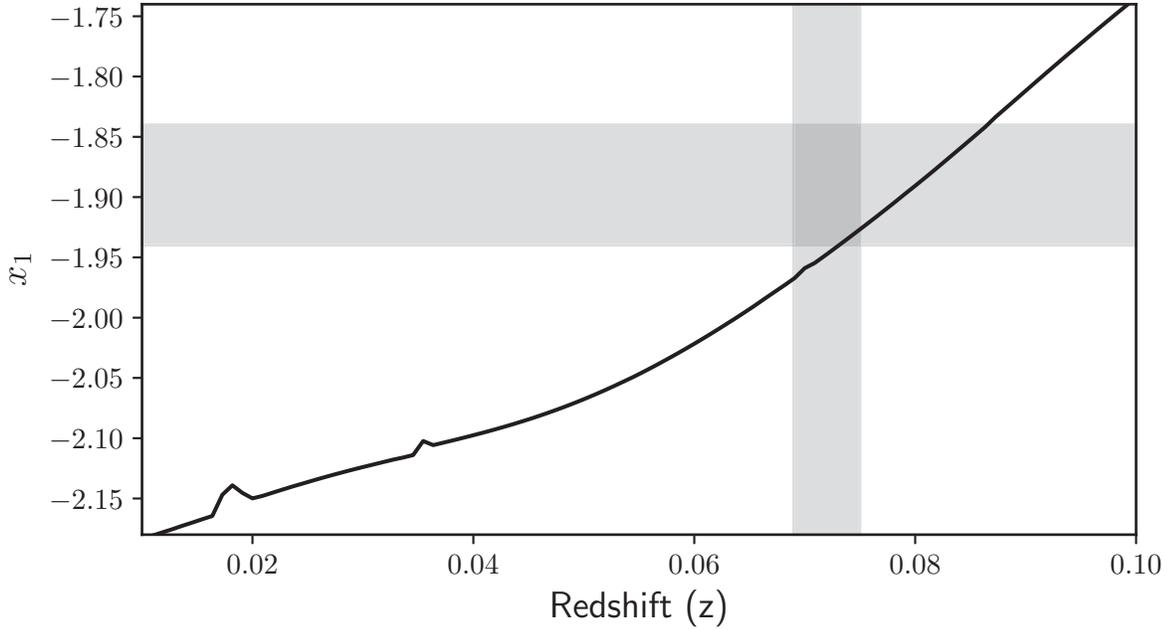}
\caption{The solid line shows the distribution of the best-fit stretch 
parameter \x1\ from the SALT2 fitting of the $VI$-band light curves of \kspn\ 
along the input redshift range of 0.01--0.1. 
The vertically shaded area represents $z$ = 0.072 $\pm$ 0.003 
which is the redshift of the source based on our \snp\ analysis. 
The horizontally shaded area represents the range \x1\ = $-$1.89 $\pm$ 0.05 
which is from the known conversion between \sbv\ and \x1\ \citep{buret14} 
for \sbv\ = 0.745 $\pm$ 0.005 that we obtain in Figure~\ref{fig:LCzfitVi} (bottom panel)
from the \snp\ $VI$-band light curve fitting.
}
\label{fig:SLAT2}
\end{figure}

The observed characteristics and measured parameters 
of \kspn\ identify the source as a \rapd\ \sni\ of transitional nature
that is more similar to Branch Normal than 91bg-like as follows.
The \rapd\ nature of the source is easily confirmed in 
Figure~\ref{fig:phil} (left panel) where we compare the $B$-band peak 
absolute magnitude and \sbv\ of \kspn\ (filled yellow star)
with those of the group of \snia\ (crosses) from \citet{buret18}.
The number of SNe gradually decreases as the \sbv\ 
values decrease (or the peak $B$-band luminosities become smaller) in the figure.
After the location of \kspn\ at \sbv\ $\simeq$ 0.54,
the gradual distribution of the SNe identified with larger \sbv\ values
disappears, and there exists only a small number of SNe as
\sbv\ decreases in this range.
The right panel of Figure~\ref{fig:phil}, 
where \dm15\ is used instead of \sbv\ for the post-peak decline rate
for the same \snia\ in the left panel,
shows the mean \dm15\ values of the 
four major subtypes of \snia\ \citep[see][]{pfp14}:
91T-like (filled red circle),
Core-Normal (filled blue circle), Broad-Line (filled orange circle), 
and 91bg-like (filled green circle) 
as well as that (\dm15\ $\simeq$ 1.62 mag) of \kspn\ (yellow star).
Note that \snia\ in the 91T-like group are slowly-evolving and overluminous,
whereas Core-Normal and Broad-Line constitute Branch Normal. 
\kspn\ clearly appears to be associated with a small number of \snia\ 
bridging the gap between Branch Normal and 91bg-like in the figure, 
showing its transitional nature. 

\moon{In \citet{buret14}, \snia\ with \sbv\ $\simlt$ 0.5 show an $I$-band peak 
$\sim$3--4 days later than $B$-band peak, while those with \sbv\ $\simgt$ 0.7 do
$\sim$2--3 days earlier. The $I$-band peak of \kspn\ precedes that of the $B$ band by
about 1.5 days, which, together with its \sbv\ $\simeq$ 0.54, places
the source in the gap connecting the two populations.
The relative strength of the secondary $I$-band peak is 
$<I>_{20-40}$ = 0.309 $\pm$ 0.002 for \kspn\ (\S\ref{subsec:lc}). 
The estimated values of \sbv\ and $<I>_{20-40}$ of \kspn\ 
follow the overall correlation between the two parameters compiled for \snia\ in the paper.
The value $\simeq$ 0.309 is low for Branch Normal but is high for 91bg-like \citep{kriet01,buret14,sriet17}, 
consistent with the interpretation that \kspn\ is of transitional nature. 
Comparison of the source with other \snia\ in terms of \sbv, $<I>_{20-40}$ and the time
delay between $B$- and $I$-band peaks shows that it overlaps with "Cool" (CL) objects
\citep[][see Figure~6 therein]{buret14} which are mostly rapid decliners, 
but close to the boundary to the group of slow decliners, 
supporting the rapidly-evolving and transitional nature of \kspn.}

The \bv\ color of \kspn\ at the peak epoch is $\simeq$ 0.08 mag, 
and this is consistent with what have been found in transitional 
\snia\ \citep{galet18}.
While the source is transitional, 
the measured values of \dm15\ and \sbv, the presence of the 
secondary $I$-band peak, the overall color evolution, 
and the results of the template fitting all show
that it is more similar to Branch Normal than 91bg-like.
According to \citet{dhaet17}, fast-declining \snia\ more luminous than
5 $\times$ 10$^{42}$ \ergs\ make a smooth connection to normal \snia\ 
as we find in \kspn\ whose peak bolometric luminosity is
(9.0 $\pm$ 0.3) $\times$ 10$^{42}$ \ergs\ (Table~\ref{tab:par} and see \S\ref{sec:bol}).

\section{Bolometric Light Curve and \ni56\ Distribution}\label{sec:bol}

The radioactive decay of \ni56\ followed by that of \co56\ 
drives the light curves of \snia\ after explosion. 
We compare here the bolometric light curve of 
\kspn\ with those expected from two different types of model \ni56\ distribution
-- centrally concentrated and stratified -- in order to investigate
how \ni56\ was distributed in the progenitor of the source as well 
as its explosion parameters of ejecta mass and kinetic energy.
During the photospheric phase of a \sni\ when the expansion is 
driven by homologous spherical shocks within the first $\sim$ 30 
days after the explosion,
the majority of its emission falls within 
the UVOIR (ultraviolet-optical-infrared) waveband \citep[][]{conet00}.
We integrate the best-fit \sni\ template of \kspn\ that we obtained
from the \snp\ template fitting (\S\ref{subsec:temp})
over the rest-frame UVOIR waveband, or $\lambda$ = 3075--23763 \AA, 
to construct its bolometric light curve.\footnote{\moon{We note that a bolometric light curve obtained this way is sometimes 
called quasi-bolometric light curve \citep[e.g.][]{bet17} given the limited wavelength
range over which the luminosity is calculated. Since the difference is expected
to be very small, we will call it bolometric luminosity light curve
throughout this paper for simplicity.}}
Figure~\ref{fig:arn}  shows the bolometric light curve of \kspn\  (filled black circles),
wherein the peak bolometric luminosity 
and epoch are  (9.0 $\pm$ 0.3) $\times$ 10$^{42}$ erg s$^{-1}$ 
and --0.5 $\pm$ 0.6 days, respectively.

\subsection{Centrally Concentrated \ni56\ Distribution}\label{sec:cenni}

In the case that the distribution of \ni56\ is strongly peaked 
towards the center of the ejected mass and that
the ejecta opacity is constant during SN explosion, 
the radioactively-powered luminosity of a \sni\ 
during the photospheric phase can be described as \citep{arn82,arn96,valet08} 
\begin{equation}
    L(x) = M_{\rm Ni} \; e^{-x^{2}} \times [(\epsilon_{\rm Ni}-\epsilon_{\rm Co}) \; C(x) + \epsilon_{\rm Co} \; D(x)]
\label{eq:lum}
\end{equation}
\noindent
where $M_{\rm Ni}$ is the total mass of \ni56.
The parameter $x$ represents a scaled time dimension of SN explosion as  
\begin{equation}
    x = \frac{t}{\tau_m}\ \ , \ \ \ \ \tau_m = \left(\frac{\kappa }{\beta_{\rm A} c}\right)^{1/2}\left(\frac{6 M_{\rm ej}^3}{5 E_{\rm ej}}\right)^{1/4}
\label{eq:time}
\end{equation}
\noindent
where $t$ is time since explosion, $\tau_m$ is the geometric mean of diffusion and expansion time scales, 
$\kappa$ is the opacity, $\beta_{\rm A}=13.8$ is a model constant for SN density distribution, 
$c$ is the speed of light, and \mej\ and \eej\ are mass and kinetic energy of the ejecta, respectively.
We adopt $\kappa$ = 0.1 cm$^2$~g$^{-1}$ dominated by the line transitions 
of \ni56\ during the photospheric phase \citep{pe00, pn14}.
In Eqn.~\ref{eq:lum}, $\epsilon_{\rm Ni} = 3.90\times10^{10}$ erg~s$^{-1}$~g$^{-1}$ and
$\epsilon_{\rm Co} = 6.78\times10^{9}$ erg~s$^{-1}$~g$^{-1}$
are the energy production rates per gram of \ni56\ and \co56, respectively.
The term $C$($x$) is related to the luminosity produced by the nuclear reaction dacay of \ni56,
and so is $D$($x$) but to \co56. 
The equation shows that the luminosity of a \sni\ in this model is mainly determined by two parameters:
the mass of \ni56\ ($M_{\rm Ni}$) and the mean time scale ($\tau_m$). 

\begin{figure}[ht!]
\includegraphics{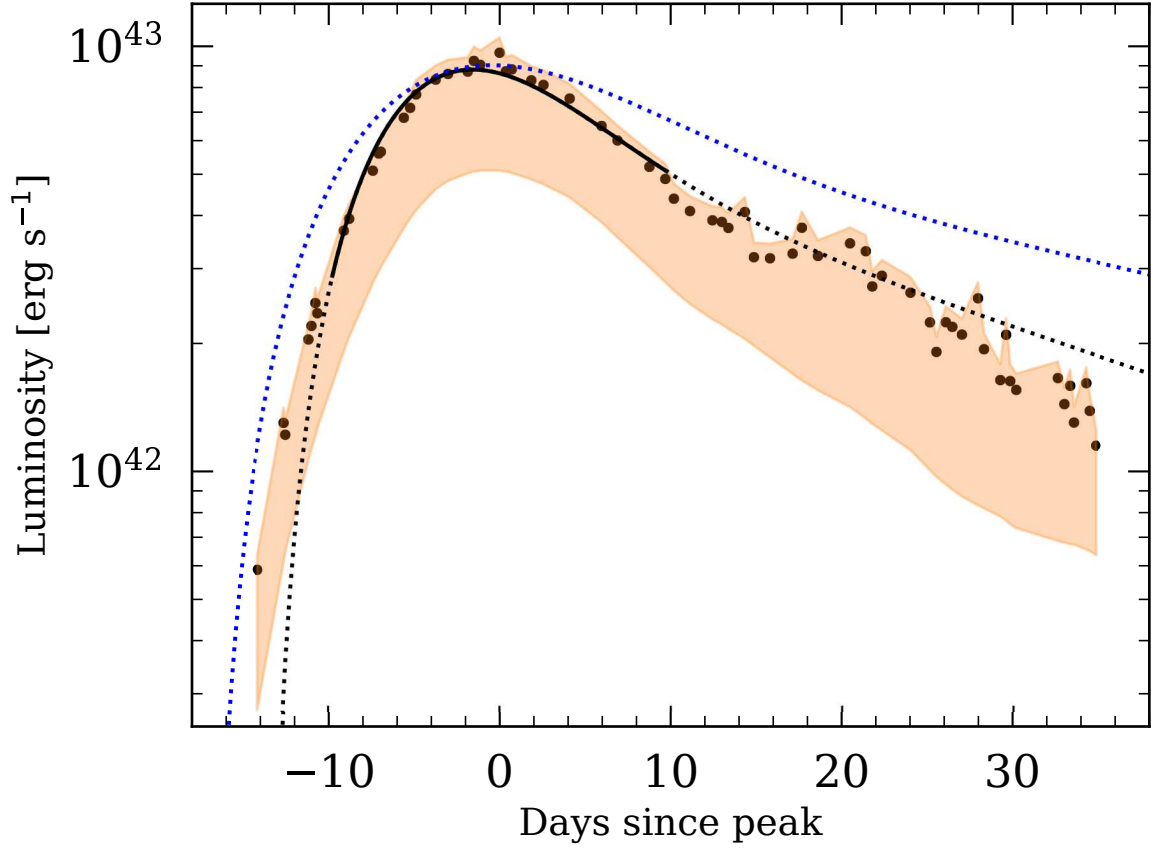}
\caption{Bolometric light curve (filled black circles) 
of \kspn\ obtained by integrating the best-fit template
over the wavelength range of 3075--23763 \AA.
The pumpkin shade represents the area between the upper and lower 
limits of the estimated bolometric luminosities. 
The upper limits are calculated by taking 
the uncertainties in photometric measurements and redshift into account,
whereas the lower limits are bolometric luminosities
integrated over the isophotal wavelength range of the $BVI$ bands only.
The abscissa is days from the epoch of the peak $B$-band
brightness determined by \snp\ fitting in the source rest frame.
The blue dotted line is for the bolometric luminosities 
calculated by Eqn.~\ref{eq:lum} 
for the centrally-concentrated \ni56\ distribution
when onset of the model and the epoch of the peak brightness 
are fixed to be the epoch of the first light (= \t0) estimated 
in the power-law fitting of early light curves (\S\ref{subsec:lc}) 
and that of the peak bolometric luminosity, respectively. 
The black solid line represents the predictions by Eqn.~\ref{eq:lum} 
obtained when only the bolometric light curve within $\pm$ 10 days
around the peak is used in model fitting.
The black dotted-line is the extrapolation of the black solid 
line to earlier and later epochs beyond the interval of 
the $\pm$ 10 days around the peak by the same fitted parameters. 
\label{fig:arn}}
\end{figure}

\moon{If we simply assume that the onset of the model 
is the same as that of the first light obtained from 
the power-law fit (\S\ref{subsec:lc}),
which gives 16.7 days as the time interval between the epochs of 
the first light and peak bolometric luminosity (or ``rise time''),
we obtain $M_{\rm Ni}$ = 0.44 $\pm$ 0.01 \msol\ 
and $\tau_m$ = 14.07 $\pm$ 0.67 days for the source
by applying Eqn.~\ref{eq:lum} at the bolometric peak.
This is equivalent to fixing Eqn.~\ref{eq:lum} using only the two parameters: the peak bolometric luminosity and the rise time,
regardless of the shape of the light curve.}
Figure~\ref{fig:arn} compares the model light curve (blue dotted line)
predicted by the two obtained parameters, i.e., $M_{\rm Ni}$ and $\tau_m$,
with the bolometric light curve (black circles) of \kspn,
where we can clearly identify over-prediction of the bolometric luminosity 
by the model across the peak, 
showing that this method is inappropriate.
This incompatibility is also confirmed with the inferred ejecta mass 
and kinetic energy of the source,
which are \mej\ = 1.86 $\pm$ 0.24 \msol\ and \eej\ = 1.35 $\pm$ 0.30 $\times$ 10$^{51}$ erg, respectively,
obtained under the assumption of 
a typical ejecta velocity \vej\ = 11000 $\pm$ 1000 km~s$^{-1}$ \citep{Scalet19}
and $E_{\rm ej}$ = $\frac{3}{10}M_{\rm ej}\varv_{\rm ej}^2$ for a \sni.
The inferred ejecta mass and kinetic energy are 
unacceptably large for a fast-evolving \sni\ \citep{Scalet19}.
This discrepancy is mainly due to  the long rise time (16.7 days) and 
the fast post-peak decline rate (\dm15\ $\simeq$ 1.62 mag) of the source
that are largely incompatible with  the model of Eqn.~\ref{eq:lum}.
In other words, while this model may accurately estimate
the total \ni56\ mass based on the peak luminosity of \kspn\ (see below), 
it is likely insufficient to reproduce the detailed evolution of 
the ascending and/or declining phases.

In order to obtain more reliable SN explosion parameters
that match the evolution of the luminosity around the peak of \kspn,
we conduct fitting of Eqn.~\ref{eq:lum} to the 
bolometric light curve of \kspn\  (Figure~\ref{fig:arn}), 
but limiting the period  between --10 and +10 days across the peak in the fitting. 
This method, which excludes the early and late part of the light curve, 
should provide more reliable explosion parameters 
since it is less affected by the assumed \ni56\ 
distribution in Eqn.~\ref{eq:lum}
while more dependent on the bulk properties of the ejecta. 
The best-fit parameters obtained this way are
$M_{\rm Ni}$ = 0.32 $\pm$ 0.01 \msol\ and $\tau_m$ = 9.45 $\pm$ 0.52 days 
which provide a significantly improved match to the observed bolometric
light curve around the peak with \chisqr\ $\simeq$ 0.42
(Figure~\ref{fig:arn}, black solid line).
\moon{The inferred ejecta mass and kinetic energy are \mej\ = 0.84 $\pm$ 0.12 \msol\ 
and \eej\ = (0.61 $\pm$ 0.14) $\times$ 10$^{51}$ erg, respectively, 
consistent with what have been observed in other fast-evolving SNe \citep{dhaet18,Scalet19,wyget19}.}
We, therefore, adopt these values as the explosion parameters of \kspn.
In this model fit, we also note that --13.5 $\pm$ 0.4 days in the SN rest frame 
is the onset of the model light curves powered by centrally concentrated \ni56 distribution
which is approximately 3.7 days after the epoch of first light (\t0, see \S\ref{subsec:lc}).
The difference of 3.7 days implies again that the centrally concentrated \ni56\ distribution 
in Eqn.~\ref{eq:lum} is inadequate to properly model the observed early light curves of \kspn,
although it is capable of providing reliable SN explosion parameters 
when fitting is limited to the light curves around the peak.
In Figure~\ref{fig:arn}, the black dotted line shows the
extrapolated model prediction of Eqn.~\ref{eq:lum} 
for early ($<$ --10 days) and late ($>$ 10 days) epochs
using the explosion parameters obtained in the fitting above.
The best-fit model apparently underpredicts bolometric luminosities at early epochs.
We attribute this underprediction and the aforementioned difference between \t0\
and the onset of the model light curves 
to a shallow \ni56\ distribution in \kspn\ 
as we detail in \S\ref{sec:strni} below.

The ejecta mass \mej\ = 0.84 $\pm$ 0.12 \msol\ places \kspn\ 
in the group of sub-Ch-mass \snia,
consistent with the results from previous studies of \snia\ that
rapid decliners are from sub-Ch-mass explosions \citep[e.g.,][]{Scalet19}. 
Note that, 
by assuming a smaller opaicty $\kappa$ = 0.08 cm$^2$ g$^{-1}$  \citep[e.g.,][]{arn82, Li19},
we obtain ejecta mass and kinetic energy 
\mej\ = 1.05 $\pm$ 0.15 \msol\ and \eej\ = (0.76 $\pm$ 0.18) $\times$ 10$^{51}$ erg, respectively,
still within the limit of the sub-Ch-mass explosion.
Recent results from extensive radiation transport simulations also 
indicate that rapid decliners are highly unlikely to be from 
Chandrasekhar or super-Chandrasekhar mass explosions \citep{gk18}.

\subsection{Stratified \ni56\ Distribution}\label{sec:strni}

As shown above, the presence of early excess emission to what is expected 
by Eqn~\ref{eq:lum} (Figure~\ref{fig:arn}) and the difference of 3.7 days 
between the epoch of the first light \t0\ 
and the onset of the model light curves of Eqn.~\ref{eq:lum} 
indicate that the real \ni56\ distribution 
of \kspn\ is different from the simple central concentration 
assumed in \citet{arn82}.
It is conceivable, as recently suggested \citep{pn14,pm16,conet18,maget20,mm20}, 
that there exists \ni56\ distributed close to the progenitor surface responsible 
for the early excess and the difference of 3.7 days. 
We investigate this possibility below based on these models.

\subsubsection{Analytic Model}\label{subsec:analytic}
According to \citet[][PN14, hereafter]{pn14}, 
a stratified \ni56\ distribution extended more towards the surface
can adequately model the bolometric light curves of other \snia\ 
such as SN~2009ig, SN~2011fe, and SN~2012cg.
PN14 modelled the local mass fraction of \ni56\ in the ejecta 
following the spherically-symmetric logistic distribution of 
\begin{equation}
    X_{56}(x) \propto \frac{1}{1+\exp{[-\beta(x-x_{1/2})]}}
\label{eq:ni}
\end{equation}
\noindent
where $x= t/t_{\rm diff}$ is a scaled depth, measured from the surface to the center, 
in unit of diffusion time ($t_{\rm diff}$) of the ejecta optical depth, while
$\beta$ and $x_{1/2}$ are shape parameters representing 
the radial decline rate of the \ni56\ distribution 
and the scaled depth at which
the distribution reaches half maximum, respectively.
During a SN explosion, a diffusion wave travels backwards 
into the expanding ejecta and SN luminosities are determined by 
the amount of \ni56\ probed by the diffusion wave which reaches the 
center of the ejecta, or $x$ = 1, some time after the SN reaches the peak luminosity.

\begin{figure}[ht!]
\includegraphics{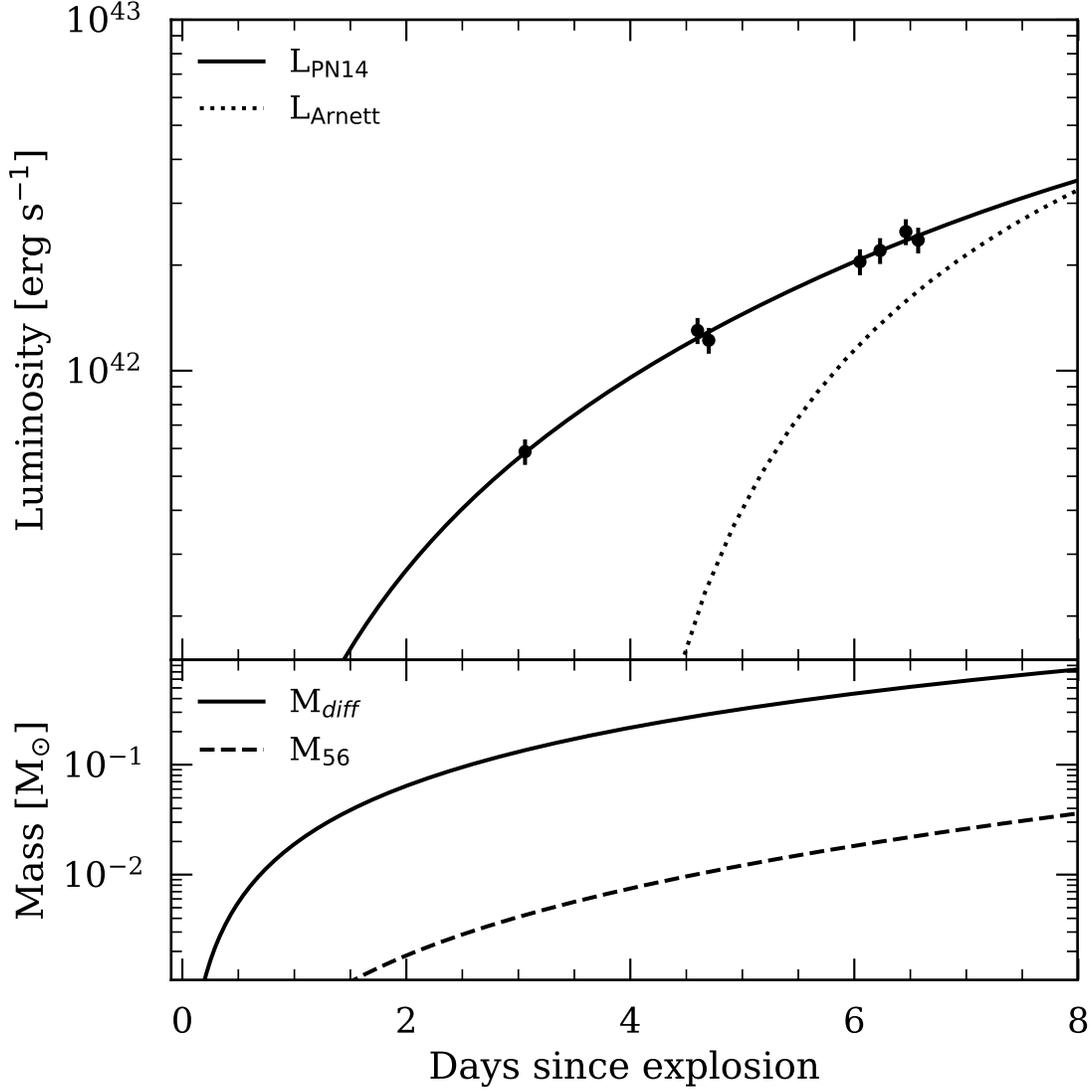}
\caption{({\it Top Panel}) Comparison between the early bolometric 
luminosity light curve (filled circles) of \kspn\ 
and the best-fit predictions (L$_{\rm PN14}$; solid line) 
from the model of 
the stratified \ni56\ distribution (Eqn.~\ref{eq:ni}) based on PN14.
The error bars represent the uncertainties measured at 68~\% confidence level.
The dotted line is the same as the dotted black line in  Figure~\ref{fig:arn},
representing the luminosities extrapolated by using
the best-fit parameters obtained from fitting the 
central part, i.e., 
within the interval of $\pm$ 10 days since peak,
of the bolometric light curve 
with the model of centrally concentrated \ni56\ distribution 
in Eqn.~\ref{eq:lum}.
The abscissa is days since the epoch of explosion,
which is the same as the epoch of the first light,
in the source rest frame.
({\it Bottom Panel}) Distribution of ejecta mass 
(M$_{\rm diff}$, solid line) and the mass of \ni56\ above the 
diffusion  wave depth (M$_{56}$, dashed line) 
calculated for the best-fit stratified \ni56\ distribution of \kspn.
}
\label{fig:SNpnmod}
\end{figure}

\begin{figure}[ht!]
\includegraphics{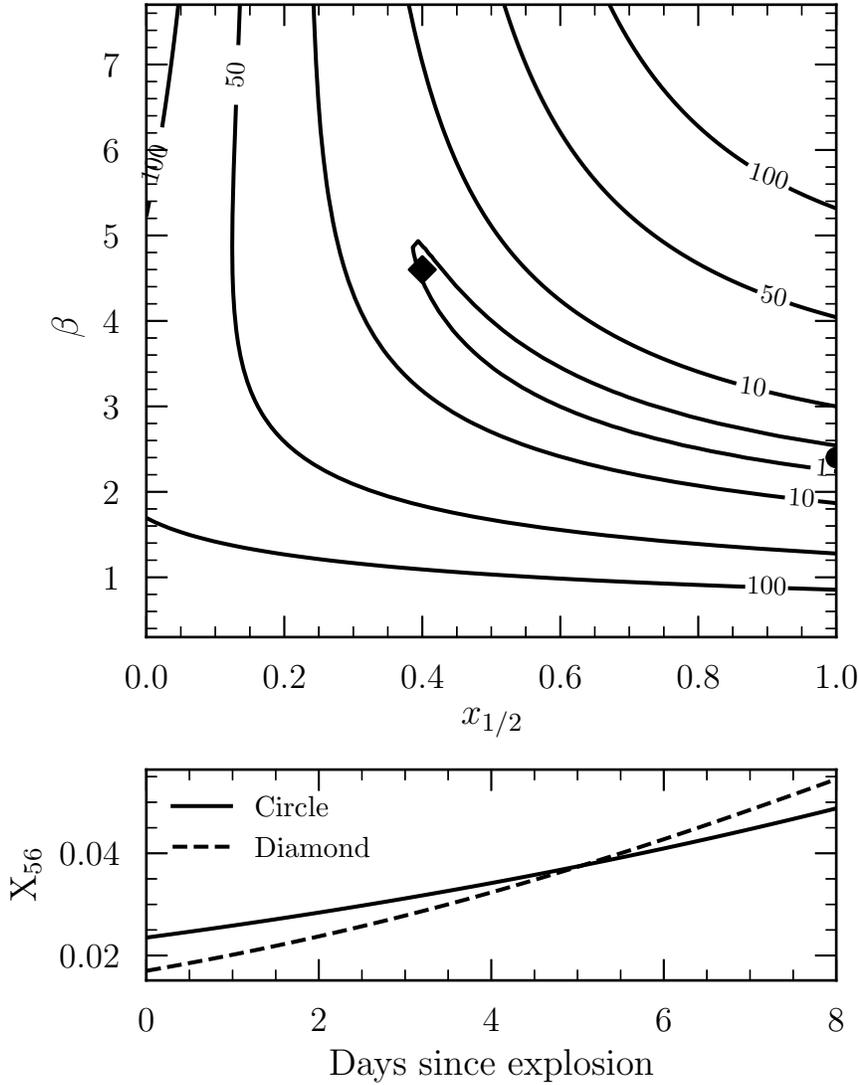}
\caption{({\it Top Panel}) Contour plot of \chisqr\ values from the fitting of
Eqn.~\ref{eq:ni} as a function of $x_{1/2}$, which is the scale depth where \ni56\
distribution is half maximum, and $\beta$, 
which is the decline rate of the distribution, in the model of PN14. 
The filled circle and diamond represent two extreme cases of
($x_{1/2}$, $\beta$) = (1.0, 2.4) and (0.4, 4.6), respectively.
({\it Bottom Panel}) The distribution of the local mass  fraction of \ni56\ 
(= X$_{56}$) for the two cases (circle and diamond) in the top panel:
solid line for ($x_{1/2}$, $\beta$) = (1.0, 2.4) and
dashed line for ($x_{1/2}$, $\beta$) = (0.4, 4.6).
The abscissa is days since the epoch of explosion,
which is the same as the epoch of the first light, in the source rest frame.
}
\label{fig:SNpnchi}
\end{figure}

Figure~\ref{fig:SNpnmod} (top panel) shows the early bolometric 
light curve (solid line) of \kspn\
predicted by the best-fit stratified \ni56\ distribution (Eqn.~\ref{eq:ni}),
confirming that the stratified \ni56\ distribution can match 
the observed early bolometric luminosities (black circles) almost perfectly
with the best-fit parameters  
$x_{1/2}$ = 1.0, $\beta$ = 2.4, and $t_{\rm diff}$ = 23.1 days 
(see Figure~\ref{fig:SNpnchi} 
below for the details of the fit).
In contrast, the extrapolation of the model light curve (dotted line)  
based on the centrally-distributed \ni56\ distribution from Eqn.~\ref{eq:lum}
shows a clear underprediction.
The bottom panel of the figure presents the early evolution 
of the estimated mass of ejecta (M$_{\rm diff}$ for solid line)
and \ni56\ (M$_{56}$ for dashed line) above the diffusion wave
(or close to the progenitor surface)
where we can identify the presence of about 0.0075 \msol\ \ni56\ mass,
which corresponds to 3.4~\% of the ejecta,
lying above the diffusion depth  at 4 days.
This tells us that a small amount of excess \ni56\ distributed shallowly
can account for the early excess emission in \kspn,
revealing that the \ni56\ distribution in the 
SN ejecta is likely more stratified towards the outer layers.

The distribution of \chisqr\ values from our model fitting 
of the stratified \ni56\ distribution, 
which is presented in Figure~\ref{fig:SNpnchi} 
as a function of $x_{1/2}$ and $\beta$,
shows that the model is reasonably well-fitted within a curved strip of
$x_{1/2}$ $\simeq$ 0.4--1.0 and $\beta$ $\simeq$ 2--5
where large $x_{1/2}$ values are paired with small $\beta$ values,
or vice versa.
This distribution pattern of $x_{1/2}$ and $\beta$ describes
either a more gradually extended \ni56\ distribution
that drops off closer to the center (= large $x_{1/2}$ and small $\beta$) or a more 
rapidly decaying distribution that drops off closer to the surface
(= small $x_{1/2}$ and large $\beta$),
both of which are consistent with a shallow \ni56\ distribution.
The bottom panel of the figure shows the distribution of the local mass fraction of \ni56\ 
(= $X_{56}$) in two extreme cases marked in the top panel --
i.e., circle for $x_{1/2}$ = 1.0 and $\beta$ = 2.4; 
diamond for $x_{1/2}$ = 0.4 and $\beta$ = 4.6 -- 
where we can confirm that both distributions require nearly the same \ni56\ 
fraction at $\sim$4-6 days after explosion.

\subsubsection{Radiative Transfer Model}\label{subsec:radiative}

\begin{figure}[ht!]
\includegraphics{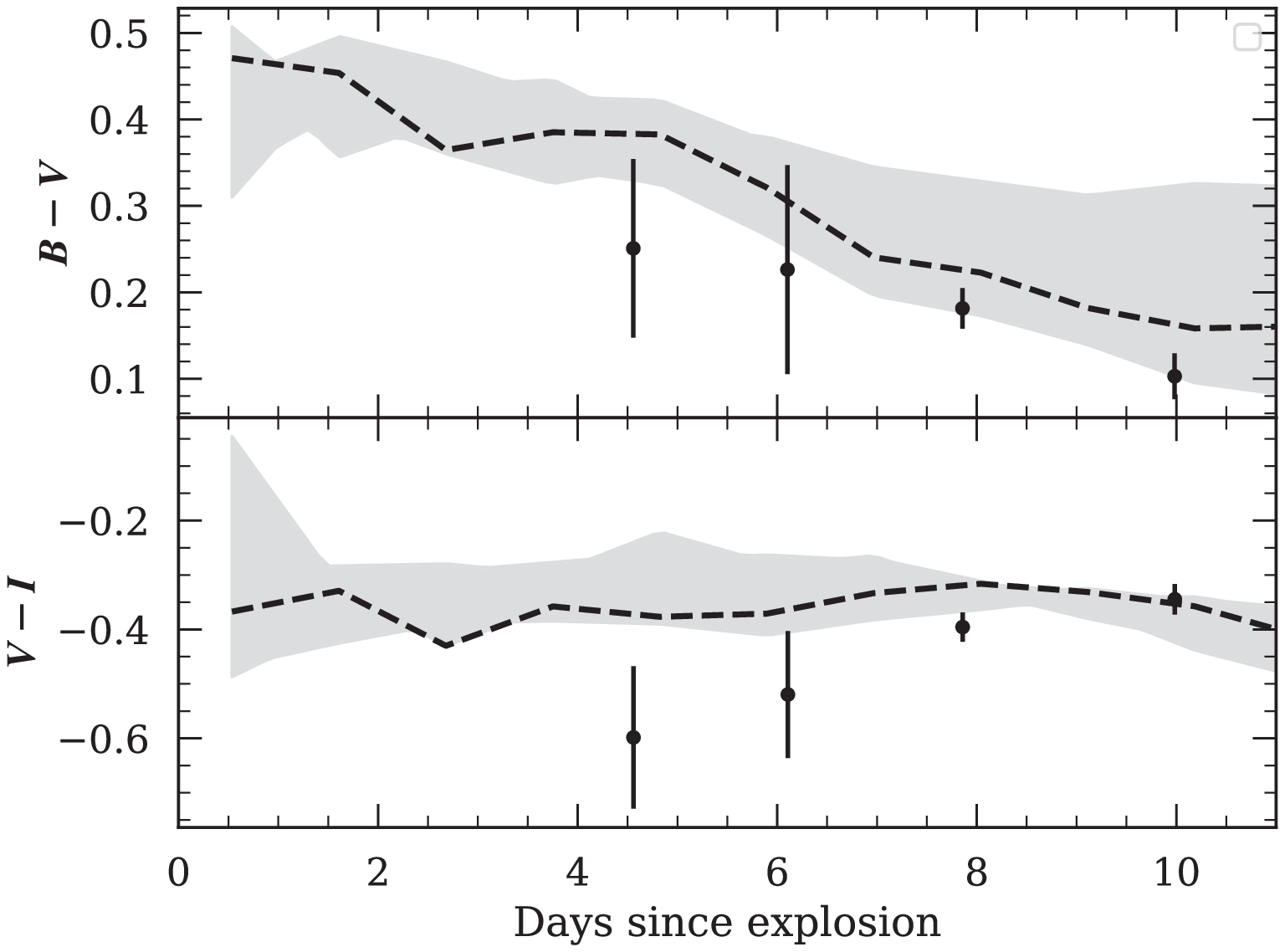}
\caption{Comparison of the \bv\ (top) and \vi\ (bottom) color evolution 
of \kspn\ (black circles), binned to 1 day interval to increase the S/N ratio over the
first 10 days post-explosion, 
to those predicted by the best-fit radiative transfer model (black dashed curves, see text)
of a Chandrasekhar-mass \sni\ with a logistically stratified \ni56\ 
distributions from \citet{maget20}. 
The error bars represent the uncertainties measured at 68~\% confidence level.
The shaded grey regions show the area of the predicted colors of 
the models with kinetic energies in the range of 
(0.5--2.2) $\times$ 10$^{51}$ ergs for the same \ni56\ mass and 
the distribution steepness obtained from the best fit.}
\label{fig:mm_Ni56}
\end{figure}

\moon{
\citet{maget20} recently provided radiative transfer-based 
model light curves of \snia\ with a logistic \ni56\ distribution  
using the following function
\begin{equation}
    X_{56}(m) = \frac{1}{1+\exp{[-s \; (m-M_{\rm Ni})/ {\rm M}_\odot ]}}
\label{eq:ni_mm}
\end{equation}
\noindent
for \ni56\ distribution, which is very similar to what is adopted in PN14 
(or Eqn.~\ref{eq:ni} in \S~\ref{subsec:analytic}). 
In this distribution, $m$ is the mass coordinate from the ejecta surface,
$M_{\rm Ni}$ is the total \ni56\ mass, 
and $s$ describes how fast the \ni56\ distribution declines. 
We compare in Figure~\ref{fig:mm_Ni56} the observed colors (black circles) of \kspn,
which are binned to 1 day interval to increase S/N ratios,
during the first 10 days post-explosion to
the best-fit radiative transfer models (black dashed curves,
$M_{\rm Ni}$ = 0.4 \msol, $s$ = 4.4, and $E_{\rm ej}$ = 0.78 $\times$ 10$^{51}$ ergs)
of \snia\ from \citet{maget20} with Chandrasekhar-mass ejecta and exponential density distributions.
The shaded grey regions in the figure represent 
the range of predicted colors for a set of radiative transfer models 
with kinetic energies  in the range of (0.5--2.2) $\times$ 10$^{51}$ ergs,
while fixing $M_{\rm Ni}$ and $s$ to those from the best-fit, 
which are 0.4 \msol\ and 4.4, respectively.
We identify in the figure that the observed early color evolution of \kspn\
is largely consistent with radiative transfer-based predictions of 
a centrally concentrated and monotonically stratified \ni56\ distribution.}

\begin{figure}[ht!]
\includegraphics{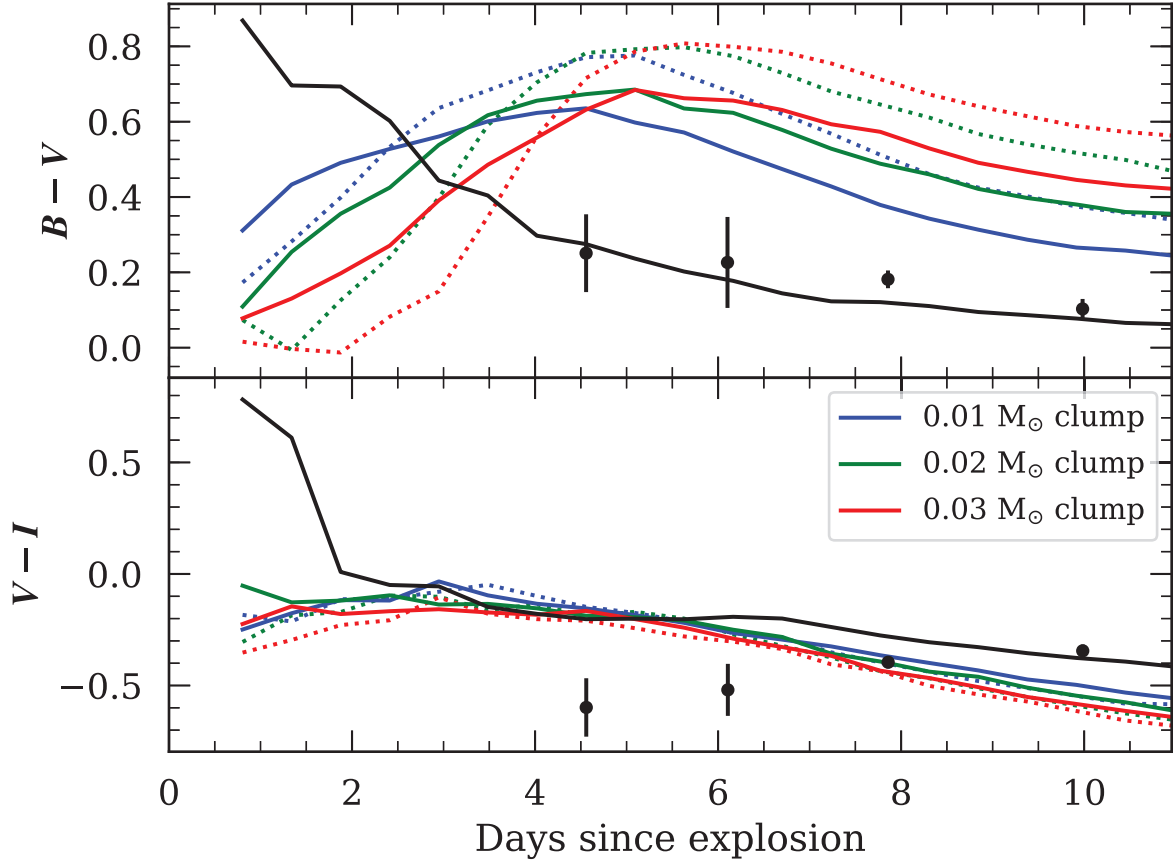}
\caption{
Same \bv\ (top) and \vi\ (bottom) colors (black circles) of \kspn\ 
as in Figure~\ref{fig:mm_Ni56}, 
but compared to the radiative transfer model light curves (colored curves) of \citet{mm20} 
for a Chandrasekhar-mass \sni\ with a logistically-stratified \ni56\ distribution
and an external shell component (see text).
The solid blue, green and red curves represent 
the cases with a shell of 0.01, 0.02 and 0.03 \msol, respectively.
The solid curves are for a shell width of 0.18 \msol, wheres the dotted curves are for 0.06 \msol.
The black solid curve represents the case without any external shell.
}
\label{fig:mm_clump}
\end{figure}

\moon{
\citet{mm20} extended the work of \citet{maget20} to compute 
radiative transfer model light curves of 
Chandrasekhar-mass \snia\ for a limited set of logistic \ni56\ distributions
with an external shell component in the outer layers of the ejecta. 
In Figure~\ref{fig:mm_clump}, we compare the observed colors of \kspn\
(as in Figure~\ref{fig:mm_Ni56}) to the predicted light curves of
the shell-added \ni56\ distributions of \citet{mm20}.
The black curve represents the best-fit model from the limited set  
of Chandrasekhar-mass \snia\ 
with  $M_{\rm Ni}$  = 0.6 \msol, $s$ = 9.7, and $E_{\rm ej}$ = 1.68 $\times$ 10$^{51}$ ergs
for the case without any shell, 
while the blue, green, red curves are for the cases of 
an added shell of 0.01, 0.02, and 0.03 \msol, respectively.
The \ni56\ distribution within these shells is assumed to be a Gaussian
centered at $m$ = 1.35 \msol\ from the center of the ejecta with 
widths of 0.18 \msol\ (solid curves) and 0.06 \msol\ (dotted curves).
As seen in the figure, the observed colors are less consistent 
with the presence of a \ni56\ shell of \simgt\ 0.01 \msol\
than the case of a centrally concentrated and monotonically 
stratified distribution of \ni56\ alone (Figure~\ref{fig:mm_Ni56}),
although we cannot rule out the possibility of a thinner \ni56\ shell of \simlt\ 0.01 \msol\  
producing colors more consistent with the observations.}

\moon{
Overall, as shown above, the observed early color evolution of \kspn\ 
within the first 10 days post-explosion is consistent with
what is expected from stratified, but still centrally concentrated, \ni56\ 
that extends to the shallow layers of the ejecta near surface. 
The color evolution is, however, largely incompatible with the presence of
a thick (\simgt\ 0.01 \msol), external shell component of \ni56\ in addition to the logistic distribution. 
This indicates that if \kspn\ originated in a \subch\ explosion triggered by 
a He-shell detonation process \citep[e.g.,][]{kroet10},
a thinner shell would have been required such as recently shown in simulations
of He-shell detonations with thin, enriched He shells \citep[e.g.][]{towet19}.}

\section{Constraint on the Progenitor of \kspn}\label{sec:pro}

The high-cadence, multi-color light curves of \kspn\ (Figure~\ref{fig:lc}) provide
a rare opportunity to place thorough constraints on the progenitors of 
rapidly-declining \snia\ of transitional nature.
For a \sni\ from a single-degenerate progenitor system composed of
a white dwarf and 
either a main-sequence (MS) subgiant or a red giant companion,
\citet[][see also \citealt{boehner17}]{kas10} calculated model
luminosities from the shock interactions between  the SN ejecta and the companion 
which are mainly determined by the mass, kinetic energy and opacity of the ejecta
as well as by the progenitor binary separation distance.
\moon{Such emission has been discussed as the source of early flashes within
roughly 5 days post-explosion in \snia\ \citep[e.g.,][]{miller18,miller20}.}
The observable luminosity from the interaction can be approximated by 
$L_{\rm int}(t) f \rm {(\theta)}$, 
where $L_{\rm int}(t)$ is the intrinsic luminosity from the interaction at time $t$
and $f \rm {(\theta)}$ = 0.982 exp[--($\theta$/99.7)$^2$] + 0.018
is the distribution of the observed luminosity as a function of the viewing angle $\theta$ \citep{ollet15}.
The maximum observable luminosity  occurs when a SN is viewed 
along the interaction axis from the side of the companion, or $\theta$ = 0$^\circ$. 

\begin{figure}
\includegraphics{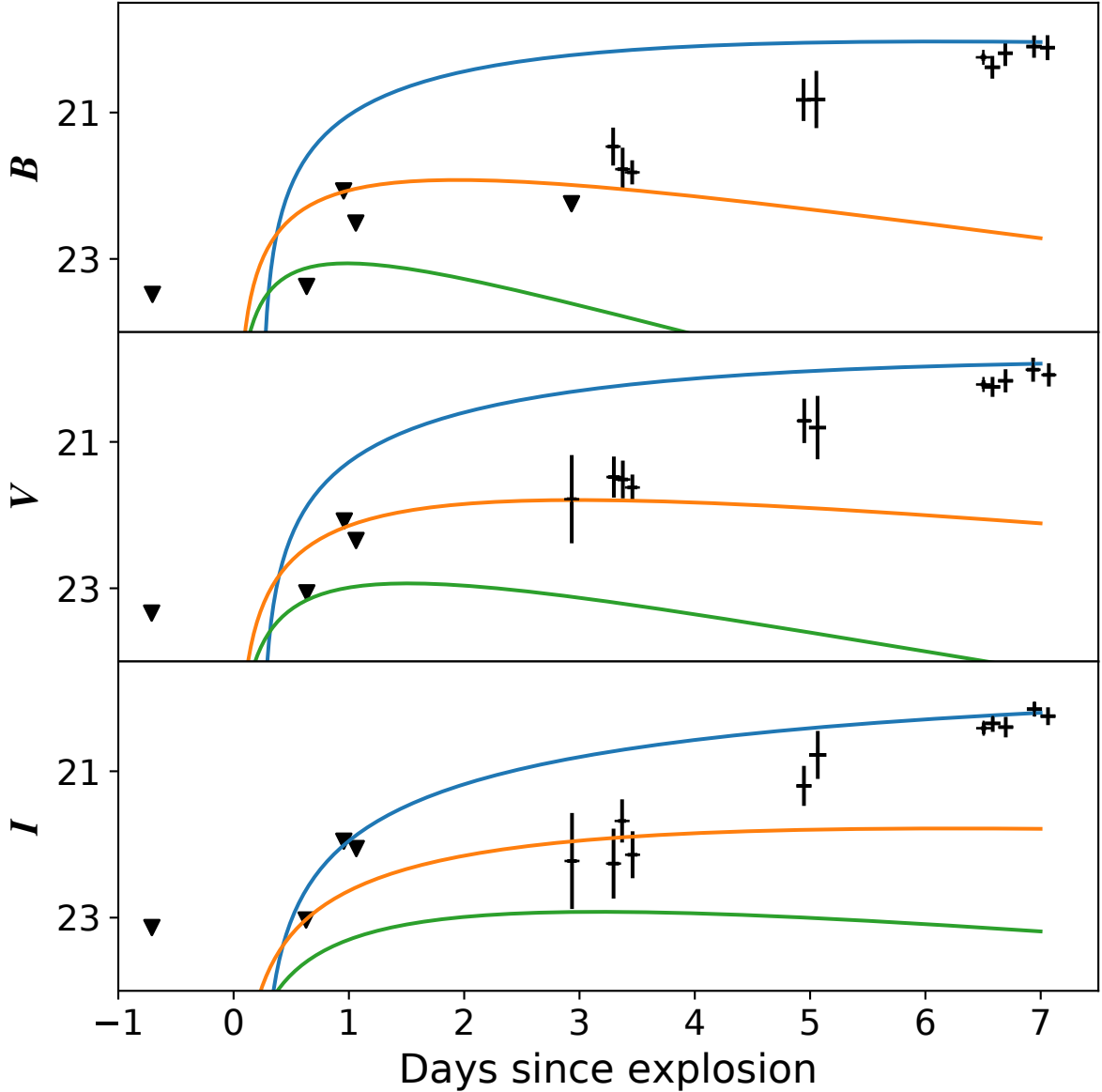}
\caption{
Comparison between the observed $BVI$ magnitudes of \kspn\ with those
expected by \citet{kas10} for the interaction between the ejecta and
the companion when viewed along the interaction axis from the companion side.
The abscissa is days since the epoch of explosion,
which is the same as the epoch of the first light (= \t0) in the source rest frame;
the ordinates are $BVI$ magnitudes from the top panel to the bottom.
The black crosses represent the observed magnitudes with the vertical
extensions corresponding to the uncertainties measured at 95~\% confidence level.
In order to increase the depth of the measurements,
we combine about five nearby individual exposures distributed within 
the time span of 2 $\pm$ 1 hours.
The black inverted triangles are detection limits at S/N = 2
which ranges between 22$\rm ^{nd}$ and 24$\rm ^{th}$ magnitude
depending on the quality of the binned images. 
The three solid lines represent model expectations 
for the 1RG (blue), 6MS (orange) and 2MS (green) cases by \citet{kas10}.
}
\label{fig:kas}
\end{figure}

Figure~\ref{fig:kas} compares the observed $BVI$ magnitudes 
(black crosses) as well as the limiting magnitudes 
(black inverted triangles) of the \kspn\ light curves 
from its early phase
with those expected by the ejecta and companion 
interaction model \citep{kas10} for three
particular companion cases of
1RG (blue solid line),  6MS (orange) and  2MS (green) 
when viewed along the interaction axis from the companion side.
These three cases are for progenitor systems where 
the Roche Lobe-filling companion is 
a red giant of 1 \msol\ (1RG),
a MS subgiant of 6 \msol\ (6MS)
and 2 \msol (2MS) located at 
2 $\times$ 10$^{13}$ cm,
2 $\times$ 10$^{12}$ cm and
5 $\times$ 10$^{11}$ cm, respectively, from the white dwarf.
We use the estimated parameters of \kspn\ in Table~\ref{tab:par}
and ejecta opacity $\kappa$ = 0.2 cm$^2$~g$^{-1}$ attributed to 
electron scattering \citep{kas10} to estimate luminosities 
from the interaction ($L_{\rm int}$) between the ejecta and companion.
As in the figure, the observed $BVI$-band brightnesses 
(including the upper limits) of \kspn\ are lower than
what are predicted by the interactions between 
the companion and ejecta 
in the 1RG case (blue line) in most of the observed epochs, 
and this is also true for the case of 6MS (orange line) at the epochs 
earlier than day 3.
This comparison shows that both the 1RG and 6MS models are 
incompatible with the observations
since the observed fluxes need to be larger than 
the model predictions to allow
for the presence of emission from the ejecta-companion interactions.
The 2MS case (green line) is different from the 1RG and 6MS
cases because only the $B$-band brightness obtained around 0.6 day 
is lower than the model prediction while 
all the other observed brightnesses (including the upper limits)
are higher.
The 2MS case, therefore, still appears to be incompatible
with the model prediction,
but with less confidence than the 1RG and 6MS cases. 
In conclusion, if the interactions between the ejecta and 
the binary companions are 
indeed viewed along the interaction axis from the companion side in \kspn,
our comparisons show that the companion of the source
was located in closer proximity to the white dwarf 
than the 1RG and 6MS cases,
and likely than the 2MS case too,
indicating that the size of the progenitor companion 
of \kspn\ is smaller than those of these three stars.
We now provide a much more thorough and general investigation
into the presence of potential emission originating from the
ejecta-companion interactions and 
conclude that it is highly likely that the companion of \kspn\
was a white dwarf, supporting the double-degenerate scenario.

\begin{figure}
\includegraphics{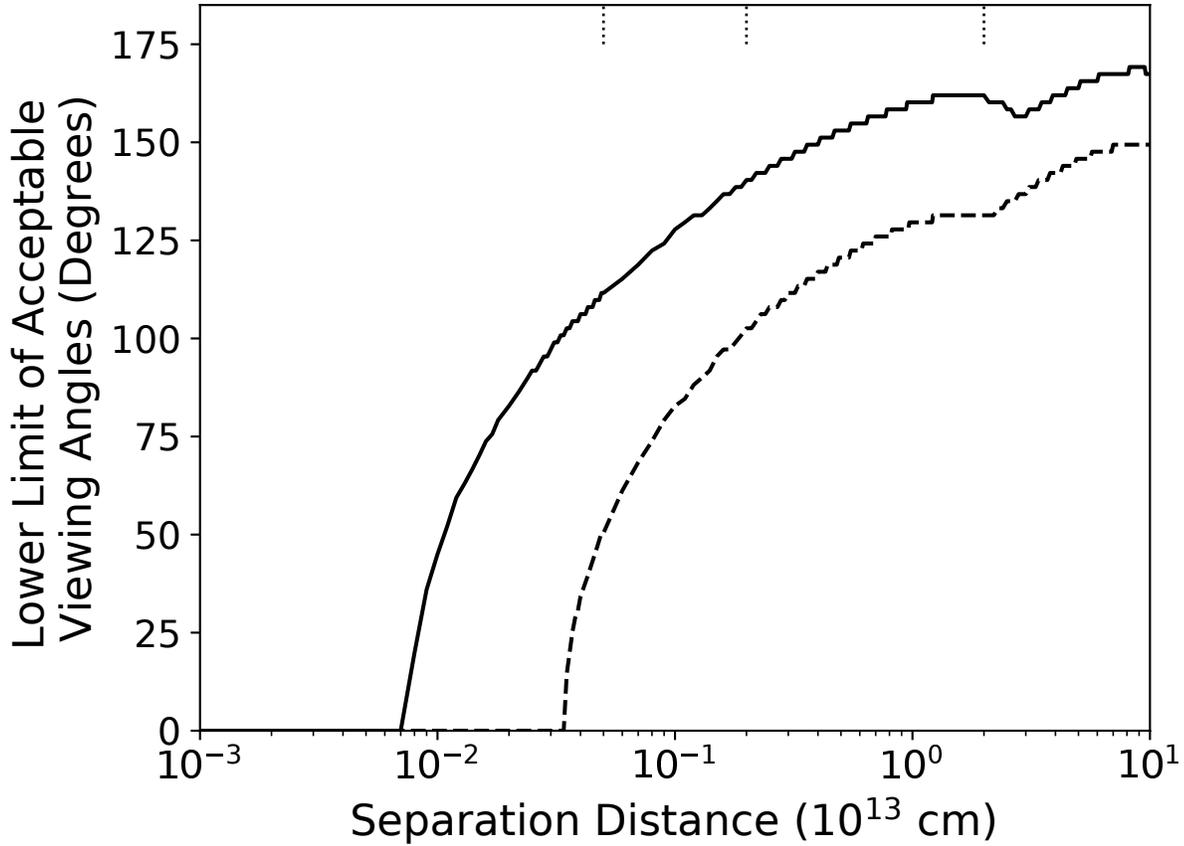}
\caption{Lower limit of acceptable viewing angles for the interaction
between \kspn\ and its potential companion predicted by \citet{kas10} as a function
of separation distance, i.e., Roche-radius separation, in unit of 10$^{13}$ cm
for two confidence levels of 68~\%\ (solid line) and 95~\%\ (dashed line) 
obtained by taking the photometric measurement uncertainties into account statistically (see text).
The three dotted vertical lines in the upper y-axis represent the separation distances
for the 2MS, 6MS and 1RG cases at 0.05 $\times$ 10$^{13}$~cm, 
0.2 $\times$ 10$^{13}$~cm and 2 $\times$ 10$^{13}$~cm, respectively.
The non-smooth feature at large separation distances
is a result of uneven quality of our data obtained at different epochs. 
}
\label{fig:kassep}
\end{figure}

We first expand our search for the signal from the potential 
interactions between the ejecta and companion in \kspn\ 
by including an extensive set of companion types,
far beyond the 1RG, 2MS and 6MS cases, 
for all possible oblique viewing angles,
i.e., $\theta$ $>$ 0\degr, 
and also by fully accounting for the uncertainties in our photometric measurements.
For this, we choose a set of 60 distances 
in the range of  (0.001--10) $\times$ 10$^{13}$~cm 
in logarithmic scale as trial binary separation distances of its progenitor.
This range of the separation distances corresponds to that of the Roche-radius separations of stars 
spanning from the smallest red dwarfs to the largest supergiants. 
We then calculate predicted brightnesses from the 
ejecta-companion interaction in $BVI$ with these 
separation distances using the model of \citet{kas10}
for a set of 100 viewing angles equally sampled 
in the range of $\theta$ = 0\degr--180\degr.
In this process, 
we compare the predicted brightnesses to the observed light curves 
at various confidence levels by adopting a Gaussian distribution 
of the observed $BVI$ magnitudes
(including the upper limits, Figure~\ref{fig:kas}) of the source
with their photometric uncertainties. 
Figure~\ref{fig:kassep} shows our results
where the abscissa is the binary separation distance
and the ordinate represents
the lower limit of acceptable viewing angles 
for the interaction between the ejecta and its companion in \kspn\ based 
on the model of \citet{kas10}
at the confidence level of 68~\%\ (solid curve) and 95~\%\ (dashed curve).
At the 95~\%\ confidence level (dashed curve), 
as in the figure,
only highly-oblique viewing angles $\theta$ $\simgt$ 130\degr\ are allowed
for the separation distances larger than 2 $\times$ 10$^{13}$~cm,
whereas all of the small viewing angles are ruled out.
For the small separation distances of $<$ 0.03 $\times$ 10$^{13}$~cm,
on the other hand, we cannot rule out any viewing angles.
In the separation distance range of 
(0.03--2) $\times$ 10$^{13}$~cm, 
the lower limit of acceptable viewing angles increases along the
separation distance, ruling out more viewing angles for larger separation distances. 
At the 68~\% confidence level (solid curve), 
they are naturally more constrained
than the 95~\% case (dashed curve) --
almost all viewing angles are excluded for large separation distances,
while we cannot rule out any angle only for the separation distances
smaller than 0.007 $\times$ 10$^{13}$~cm.

\begin{figure}[ht!]
\includegraphics{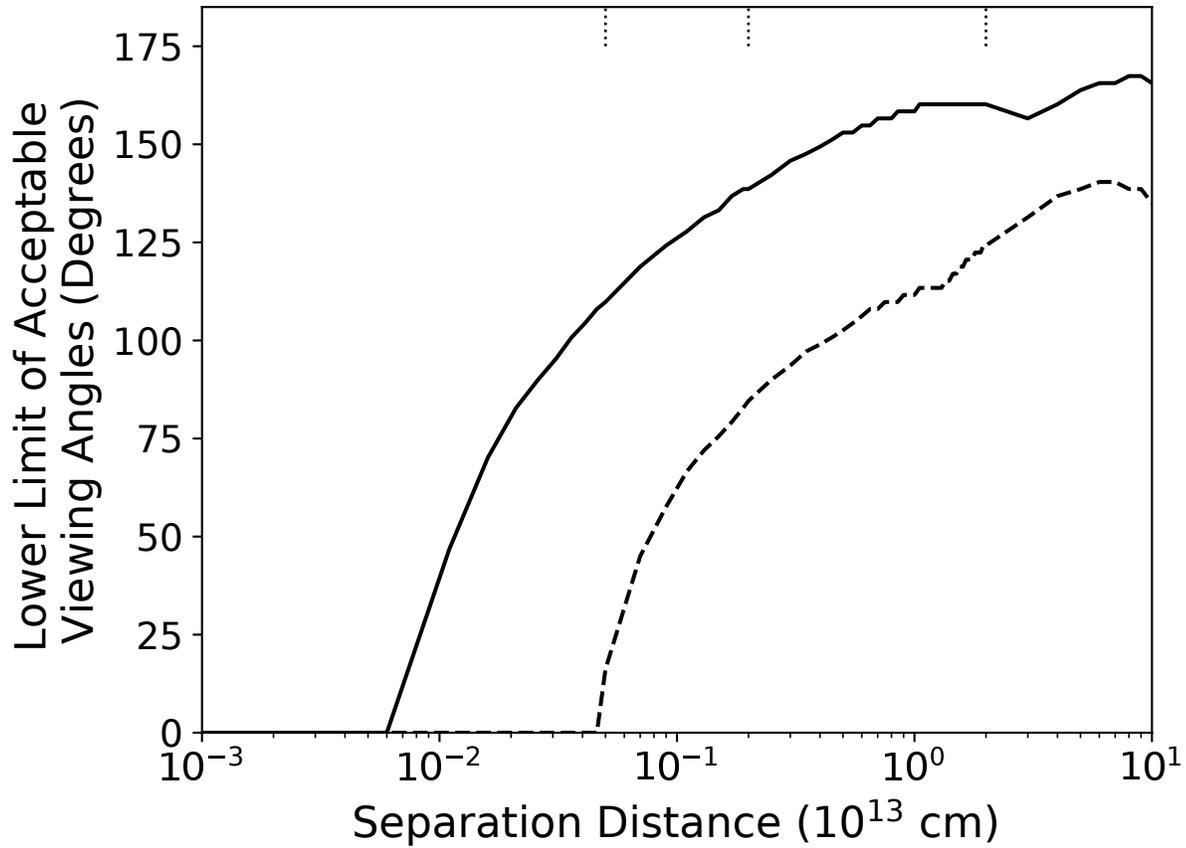}
\caption{
Same as Figure~\ref{fig:kassep}, but for the confidence levels
(solid line for 68~\%; dashed line for 95~\%)
obtained from 40000 Monte Carlo simulations of the model luminosities 
from ejecta-companion interactions \citep{kas10}. The luminosities are
computed by randomly selecting the values of a comprehensive set of the
observed and estimate parameters (including redshift, ejecta mass and kinetic
energy, epoch of the first light as well as photometric measurements)
of \kspn\ under the assumption that their uncertainty distributions are Gaussian.
\label{fig:kasepo}}
\end{figure}

The results shown in Figure~\ref{fig:kassep} accommodate the 
photometric uncertainties of the light curves of \kspn, 
but not those of the estimated values of the redshift, epoch of the first light, and ejecta mass and kinetic energy
needed to compute the model luminosities from the ejecta-companion interactions.
In order to reflect the uncertainties of those parameters 
in our analyses, we conduct 40000 
Monte Carlo simulations of the light curves of the ejecta-companion 
interactions predicted by \citet{kas10}
by randomly selecting the values of these parameters 
under the assumption that they follow a Gaussian distribution 
with the measured uncertainties.
We use the same set of the separation distances 
and viewing angles that 
we used above for Figure~\ref{fig:kassep}
and then conduct comparison between the observed light curves 
of \kspn\ with the model light curves. 
Figure~\ref{fig:kasepo} shows the results of our comparison,
wherein all the viewing angles are allowed for the separation distances 
up to $\simeq$ 0.045 $\times$ 10$^{13}$ cm at the confidence level of 95~\%
when all the uncertainties are statistically accounted for. 
The separation distance of 0.045 $\times$ 10$^{13}$ cm
is about 90~\% of that of 2MS after which the lower limit of acceptable viewing angle increases.
At the separation distances larger than that of 1RG, 
only large viewing angles, i.e., $\theta$ $>$ 125\degr, are allowed.
The constraints on the viewing angles become more stringent at the 68~\% confidence level
with even small viewing angles being ruled out at very short separation distances.

Our analyses above apparently show  that 
there is not much chance for the companion of \kspn\ to have been a red giant
considering that the lower limit of acceptable viewing angles
rules out most of the viewing angles
for such a large star.
Note that it is highly unlikely the companion was 
at a different location from the Roche-radius separation where
it can provide the stable accretion needed to trigger the explosion.
A MS subgiant appears to be more probable 
than the red giant in our analyses, 
especially at highly-oblique viewing angles;
however, sustaining substantial accretion from such a MS companion 
for a \sni\ detonation is more challenging.
These results indicate that a double-degenerate system is much more 
likely to have been the progenitor system of \kspn,
which agrees well with the results of the statistical 
searches for ejecta-companion shock interactions
in the surveys of SDSS, SNLS and TESS \citep{hayden10,bianco11,fausnaugh19}
and also with those of individual SN studies of 
SN2011fe \citep{nuget11,liet11}, 
SN2012ht \citep{yamanaka14}, 
SN2013gy \citep{holmbo19}, 
ASASSN-14lp \citep{shapet16} 
and SN2012cg \citep[][although see \citealt{maret16}]{shapet18},
strengthening the case of small companions for the majority 
of \snia\ progenitors \citep[e.g.,][]{ollet15}. 
We, however, also note that some SNe, e.g.,
SN2014J \citep{goobar15}, 
SN2017cbv \citep{hosseinzadeh17} and 
SN2018oh \citep{shapet19,dimitriadis19}, 
do show indications of early excess emission that could come 
from ejecta-companion shock interactions, 
although it is possible that some other processes---such as circumstellar interactions or a shallow layers of \ni56---are responsible for it.

\section{Missing Host Galaxy for \kspn}\label{sec:host}

\begin{figure}[ht!]
\includegraphics[width=\textwidth]{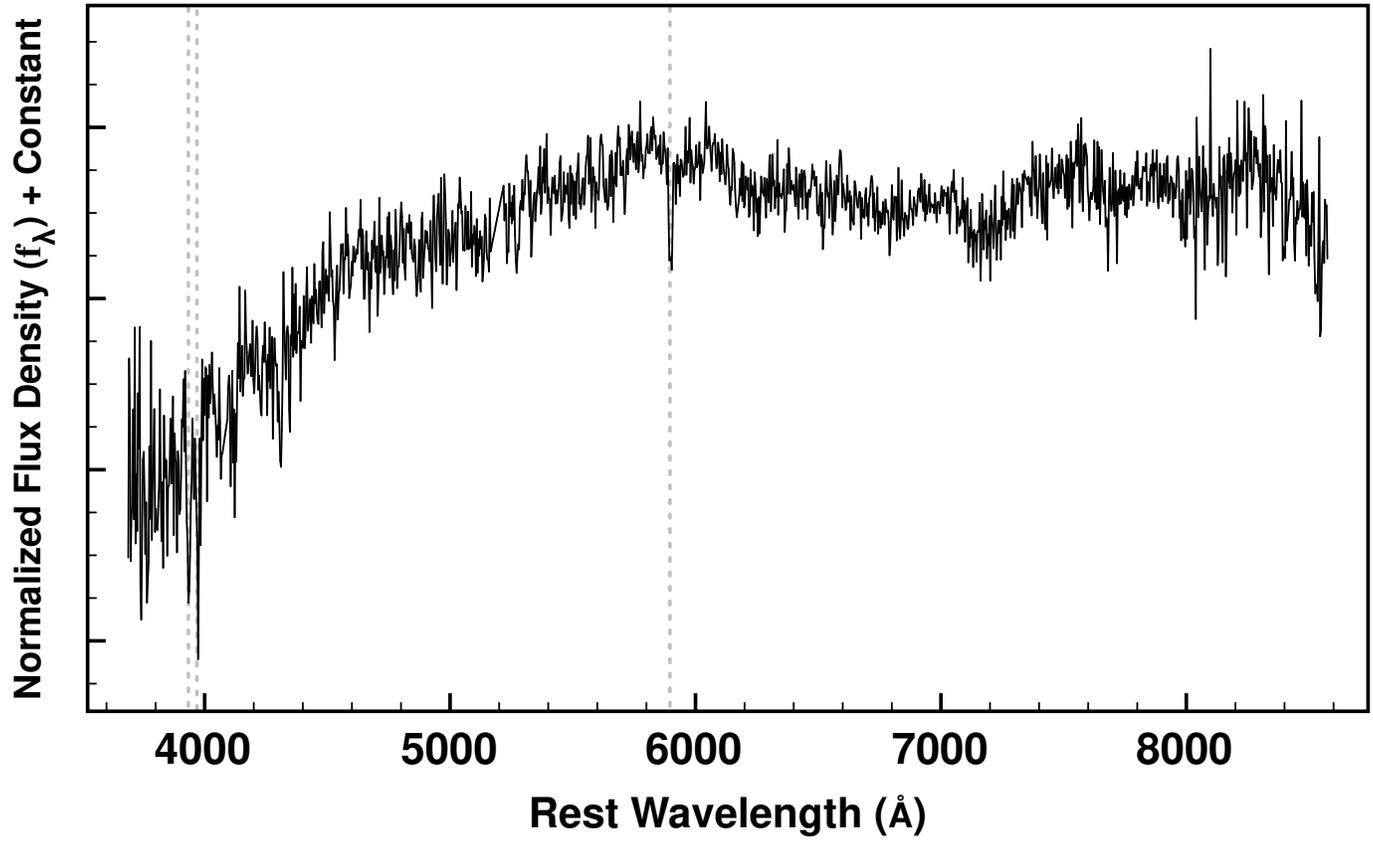}
\caption{Spectrum of the nearby galaxy \galgg\ obtained with the Magellan telescope.
Absorption lines of Ca II H+K and Na I D are marked by the dashed vertical lines.
Their wavelengths are consistent with the redshift $z$ = 0.167 $\pm$ 0.001
for the galaxy \galgg.
\label{fig:gal}}
\end{figure}

\begin{figure}
\includegraphics[width=\textwidth]{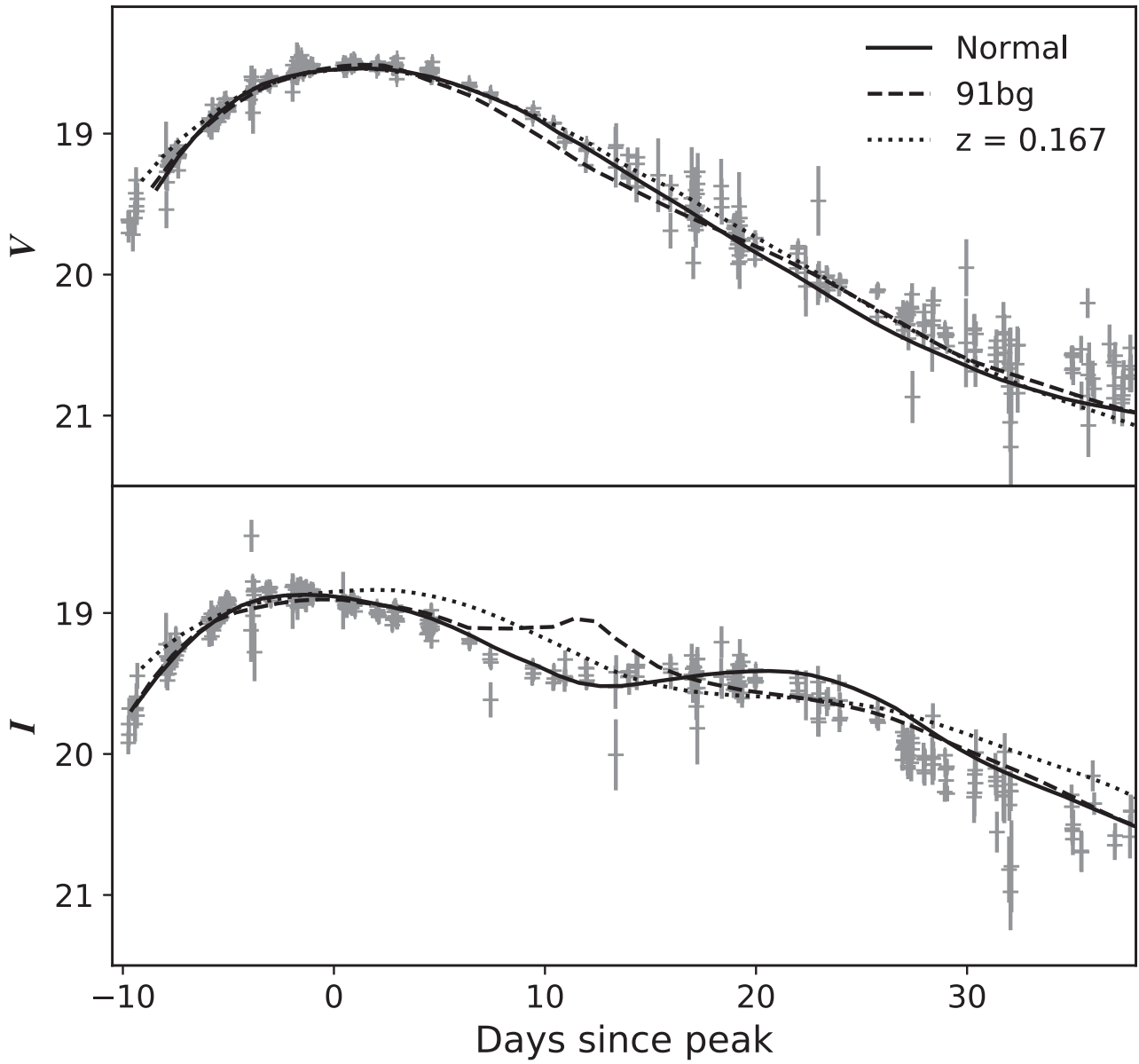}
\caption{Same as Figure~\ref{fig:LCtemplate} but with the
dotted lines representing the best \snp\ template fits
of \kspn\ with $z$ = 0.167.}
\label{fig:z0163fit}
\end{figure}

As shown in Figure~\ref{fig:det} (see also \S~\ref{sec:obs}),
no host galaxy underlying \kspn\ 
is detected in our deep stack images reaching the sensitivity
limits of $\simeq$ 27.8 ($B$), 28.5 ($V$) and 28.2 ($I$) \mpas, while the source
\galgg, which is the only apparently extended source  near \kspn, 
is located $\sim$27\arcsec\ away in the southwestern direction.
In order to understand the nature of \galgg\ and its potential 
connection to \kspn,
we fit its $V$-band surface brightness
with the S{\'e}rsic profile
$\mu$ = $\mu_0$ +1.0857 $b_n(r/r_{\rm e})^{1/n}$
(where $\mu_0$, $r$, $r_{\rm e}$ and $n$ are the central 
surface brightness, radius, effective radius and S\'ersic 
curvature index, respectively, and 
$b_n$ = 1.9992$n$--0.3271, see \citealt{graham05})
to obtain $\mu_{0}$ = 19.04 $\pm$ 0.14 \mpas,
$r_{\rm e}$ = 3\farcs63 $\pm$ 0\farcs03 and $n$ = 2.05 $\pm$ 0.06.
The apparent $V$-band magnitude and the \bv\ color of \galgg\ 
are $V$ $\simeq$ 17.38 mag and \bv\ $\simeq$ 1.4 mag, respectively.
These fitted parameters and color of \galgg\ are similar to those 
found in early-type galaxies \citep{baset13, valet11}, 
which is consistent with its spectroscopic properties.
Figure~\ref{fig:gal} shows our Magellan spectrum of \galgg\ (see \S~\ref{sec:obs} 
for the details of the observations)
with clear detections of Ca H+K, G and Na ID absorption lines, 
typical of early-type galaxies. 
We determine the redshift of \galgg\ to be 
$z$ $\simeq$ 0.167 $\pm$ 0.001
using the measured wavelengths of the absorption lines,
where the uncertainty is due to the rms wavelength 
solution error of 0.135 \AA.
The measured redshift is much larger than $z$ = 0.072 $\pm$ 0.003 inferred for \kspn\ by
fitting to \sni\ template (\S\ref{subsec:temp}).
Figure~\ref{fig:z0163fit} compares the best-fit \snp\ Branch Normal template (dotted curve) 
obtained at $z$ = 0.167 with those in Figure~\ref{fig:LCtemplate}, 
showing clearly that it gives significantly 
worse results than the fit obtained at $z$ = 0.072.
In addition, in order for \kspn\ to be a \sni\ with a high redshift $z$ = 0.167,
its observed $B$-band peak brightness requires the SN to either have a
peculiar motion greater than 14000 \kms\ or peak absolute magnitude
more luminous than --21.8 mag with \dm15\ $\simeq$ 1.96 mag.
Both of these are unrealistic since the required peculiar motion is more than
20 times greater than what can be expected for the source (see \S\ref{subsec:temp})
and the required peak luminosity and \dm15\ make the source an extremely luminous 
\sni\ with an exceptionally large decline rate.
We, therefore, conclude that \galgg\ is unrelated to \kspn\
and is an early-type galaxy located at a much larger redshift.
This leaves \kspn\ still hostless. 

\begin{figure}
\includegraphics[width=\textwidth]{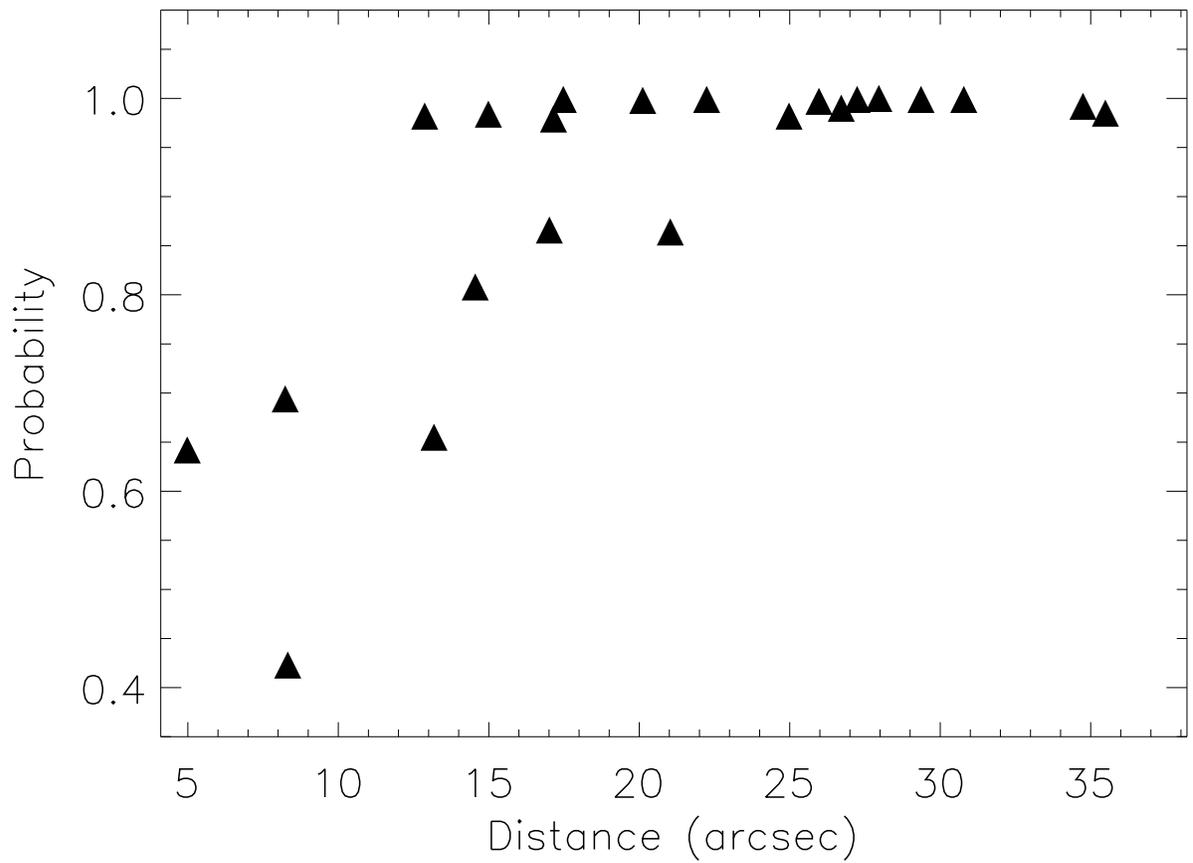}
\caption{Probability of chance coincidence of 22 closest sources
(excluding the galaxy \galgg)
identified in our deep $BVI$ images as a function of 
distance from \kspn. 
\label{fig:maria}}
\end{figure}

In order to investigate whether another nearby detected source
may be the host galaxy of \kspn, we then 
calculate probabilities that 22 sources identified within
about 35\arcsec\ from \kspn\ in our deep $BVI$ images
with S/N $>$ 2
are random galaxies coincidentally located in the field
by adopting the methodology in \citet{bkd02} using their 
magnitudes and distances from \kspn\ 
\citep[see also][and the details therein]{berger10}.
In Figure~\ref{fig:maria}, which shows 
the calculated probabilities of 
chance coincidence as a function of distance,
all the sources have high ($>$ 0.4) chance coincidence probabilities 
with the majority near 1. 
This tells us that all these sources near \kspn\
are likely to be coincident by chance
and are unlikely related to the SN.

The absence of any host galaxy candidate of \kspn\ 
in our deep stack images hints to the nature of its 
host galaxy since,
at the luminosity distance of $\simeq$ 310 Mpc (or $z$ $\simeq$ 0.072), 
regular galaxies should be easily identifiable as an extended object \citep[see][for example]{aet14} in our images.
Considering that there is no such extended object 
other than \galgg, which is at a much higher redshift, 
in the vicinity of \kspn, 
it is highly likely that the host galaxy of \kspn\ is 
a faint dwarf galaxy.
The limiting magnitudes of an unresolved source in our images 
are $B\simeq 25.83$, $V\simeq 25.30$, and $I\simeq 24.74$ mag.
This corresponds to an absolute magnitude limit of $\simeq$ --12 mag 
in the $V$ band at $z$ = 0.072 and it is 
compatible with the previously-known $V$-band absolute magnitude range
of dwarf galaxies \citep{tol09}.
The nearest unresolved source to \kspn\ in Figure~\ref{fig:det} 
is about 5\arcsec\ away
from \kspn\ with $V$- and $I$-band magnitudes of 
24.14 $\pm$ 0.15 and 23.99 $\pm$ 0.17 mag, respectively. 
If this unresolved source is a dwarf galaxy hosting \kspn,  
the SN is located $\sim$ 6.6 kpc away from the center of the host galaxy.
Its $V$-band absolute magnitude is --13.23 mag at $z$ = 0.072, 
and a dwarf galaxy with such magnitude 
is expected to have $\simeq$ 25 \mpas\ effective surface brightness
and $\simeq$ 1 kpc effective radius \citep{cb15,tol09}.
This requires that \kspn\ exploded at a location away 
from the center of a dwarf galaxy more than six times 
of its effective radius. 
According to \citet{k13}, the stellar density in a dwarf galaxy 
at such a location drops significantly more, i.e., $\gg$ 50 times, 
than the central part.
We, therefore, conclude again that it is highly unlikely that 
any source identified in our deep stack images is 
the host galaxy of \kspn\ and that the host galaxy of the SN
is most likely a dwarf galaxy fainter than our detection limit.
The inferred insignificant host galaxy extinction of \kspn\  
(\S\ref{sec:obs} and \S\ref{subsec:color}) points to 
the outskirts of its potential host galaxy as the explosion location.

Recently there has been a growing number of \snia\ detected 
in low-luminosity dwarf galaxies as well as those that remained hostless. 
These include
(1) SN1999aw, a luminous and slow-decaying SN from a 
very faint galaxy with $M_V=-12.4$ mag \citep{Strolger2002};
(2) SN2007qc from an extremely faint host galaxy with $M_B \sim -11$ mag \citep{Quimby2012};
(3) SN2007if, a luminous super-Chandrasekhar-mass \sni\
detected in a host galaxy with $M_g=-14.45$ mag \citep{Childress2011};
and (4) PTF10ops, a peculiar type SN with subluminous spectral 
properties but with a normal light-curve width 
while remaining hostless to the detection 
limit of $r$ $\gtrsim$ --12 mag \citep{maguire11}. 
It is also noteworthy that the group of Ca-rich transients, 
whose thermonuclear origin is still under debate, 
have consistently been found in the outskirts of their host galaxies 
where there is no apparent stellar population,
preferring high-velocity progenitors to low-metallicity environment \citep{yuan13,foley15,lyman16,lunnan17}.
\cite{Graham2015} identified the host of a \sni\ with 
$M_V$ $\simeq$ --8.4 mag which may be either
a dwarf galaxy or a globular cluster from a nearby elliptical 
galaxy along with two other cases where no host galaxy 
is identified to the limit of $M_R$ $>$ --9.2 mag, 
suggesting that their progenitors most likely belong to the intracluster 
stellar population.
Although \snia\ from a faint host,
including those from intra-cluster environment and dwarf galaxies, 
can potentially bring us an important insight into the connection between
their progenitors and stellar population and be used to trace missing dwarf galaxies in the 
local universe \citep[e.g.,][]{Graham2015,cb15},
our understanding of such \snia\ is still 
highly incomplete, mainly due to the lack of statistically meaningful sample size. 
The identification of \kspn\ as a rapidly-declining hostless
\sni\ of transitional nature indicates that this type of  
\snia\ can be produced
in faint dwarf galaxies and that the coupling between
specific types of \snia\ and host galaxy environment may not be
as strong as previously thought considering that
the transitional \snia\ have largely been detected
in early-type galaxies \citep[e.g.,][]{ashet16, mrbh07}.

\section{Summary and Conclusion}\label{sec:sum}

In this paper, we report the discovery and 
identification of \kspn\ 
as a rapidly-declining hostless \sni\ of transitional nature
likely originating from a sub-Ch explosion in a double-degenerate progenitor
based on high-cadence, multi-color observations made with the KMTNet.
We summarize our results and conclusion as follows.

\begin{itemize}
\item The observed light curves and colors of \kspn\ 
are compatible with a \rapd\ \sni\ at $z$ $\simeq$ 0.072
whose properties are intermediate between Branch Normal 
and 91bg-like, but much closer to the former, 
with clear signs of transitional nature.
While the evolution of its early light curves is well fitted 
with a power law representing a homologous expansion
powered by \ni56\ radioactive decay which is mediated by photon diffusion processes,
the overall \bv\ and \vi\ color evolution of \kspn\ is largely
synchronous with that of the $I$ band as found in other \snia.
We identify the presence of an early redward evolution
in the \vi\ color prior to --10 days since peak in the SN
before it enters the previously-known phase of blueward evolution.
This early redward evolution in \vi\ 
has not been much studied but may bear an important clue
to understanding the physical conditions of SN explosions.

\item The \pp\ and \csp\ of \kspn\ are 
\dm15\ = 1.62 $\pm$ 0.03 mag and \sbv\ = 0.54 $\pm$ 0.05, respectively,
which place the source in the gap between the two groups of 
Branch Normal and 91bg-like with the peak luminosity of
(9.0 $\pm$ 0.3) $\times$ 10$^{42}$ \ergs. 
The transitional nature of the source is also confirmed with 
the relative strength ($\simeq$ 0.309) 
of the secondary $I$-band peak of the source 
and its \bv\ $\simeq$ 0.08 mag color at the peak epoch.
We obtain $M_{\rm Ni}$ =  0.32 $\pm$ 0.01 \msol, 
\mej\ =  0.84 $\pm$ 0.12 \msol\ and 
\eej\ = (0.61 $\pm$ 0.14) $\times$ 10$^{51}$ erg,
which make \kspn\ a \sni\ explosion with
a sub-Ch mass.

\item The bolometric light curve of \kspn\ shows the presence of
an early excess emission to what is expected from a centrally-concentrated
\ni56\ distribution. 
\moon{We find that a stratified \ni56\ distribution 
extended more shallowly to the surface of the progenitor provides 
a good  match to the observed bolometric light curve,
while the presence of a thick, \simgt\ 0.01 \msol, external shell
is largely incompatible with the observed early color evolution.}
Thorough comparisons between the observed light curves 
and those predicted from the ejecta-companion interactions 
clearly prefer a small binary separation distance for the progenitor, 
favoring the double-degenerate scenario for its origin.

\item Even in our deep stack images reaching the sensitivity limit
$\mu_{BVI}$ $\simeq$ 28 \mpas,
\kspn\ remains hostless, suggesting that its host galaxy 
is a faint dwarf galaxy. This contradicts to what has been previously
thought for the types of host galaxies that produce transitional \snia.
It will be worthwhile to investigate the nature of the host galaxy 
of \kspn\ with deeper imaging observations than presented in this paper
that can shed new insight into the relationship between 
host galaxies and types of \snia.

\end{itemize}

\acknowledgments
This research has made use of the KMTNet facility operated by the Korea Astronomy and Space Science Institute and the data were obtained at three host sites of CTIO
in Chile, SAAO in South Africa, and SSO in Australia. 
We acknowledge with thanks the variable star observations
from the AAVSO International Database contributed by
observers worldwide and used in this research. 
DSM was supported in part by a Leading Edge Fund from the Canadian Foundation 
for Innovation (project No. 30951) and a Discovery
Grant (RGPIN-2019-06524) from the Natural Sciences and Engineering Research Council (NSERC) of Canada. 
MRD acknowledges support from NSERC through grant RGPIN-2019-06186, 
the Canada Research Chairs Program, 
the Canadian Institute for Advanced Research, 
and the Dunlap Institute at the University of Toronto. 
HSP was supported in part by the National Research Foundation of Korea (NRF) grant
funded by the Korea government (MSIT, Ministry of Science and ICT; No. NRF-2019R1F1A1058228).
JA is supported by the Stavros Niarchos Foundation (SNF) and the Hellenic Foundation for Research and Innovation (H.F.R.I.) under the 2nd Call of "Science and Society" Action Always strive for excellence Theodoros Papazoglou" (Project Number: 01431)

\vspace{5mm}
\facilities{KMTNet, Magellan, AAVSO}
\software{
Astropy \citep{astropy13,astropy18},
\snp\ \citep{buret11},
SCAMP \citep{bertin06},
SWARP \citep{bet02},
SiFTO \citep{conet08}
}

\clearpage

\clearpage

\clearpage

\clearpage

\clearpage

\clearpage

\clearpage

\clearpage

\clearpage

\clearpage

\clearpage

\clearpage

\clearpage

\clearpage

\clearpage

\clearpage

\clearpage

\clearpage

\clearpage

\begin{deluxetable}{cccc}
\tabletypesize{\footnotesize}
\tablecolumns{4} 
\tablewidth{0.99\textwidth}
 \tablecaption{Observed magnitudes of \kspn}
 \tablehead{
 \colhead{Time [MJD]}  & \colhead{Magnitude$\rm ^a$ [mag]} & \colhead{Error [mag]} & \colhead{Band}
 } 
\startdata 
57289.89826 & 21.292 & 0.368 & $B$ \\
57289.90000 & 21.447 & 0.333 & $V$ \\
57289.91440 & 21.648 & 0.487 & $B$ \\
57289.91770 & 21.587 & 0.322 & $I$ \\
57289.93130 & 20.971 & 0.283 & $B$ \\
57289.93338 & 21.332 & 0.277 & $V$ \\
57289.96603 & 21.036 & 0.297 & $B$ \\
57289.96760 & 21.195 & 0.208 & $V$ \\
57289.98403 & 21.517 & 0.355 & $V$ \\
57289.98574 & 21.267 & 0.246 & $I$ \\
57289.99826 & 21.188 & 0.320 & $B$ \\
57290.00235 & 21.652 & 0.308 & $I$ \\
\enddata
\tablenotetext{{\rm a}}{The $B$- and $V$-band magnitudes are in the Vega system,
while the $I$-band magnitudes are in the AB system (see text).}
\tablecomments{Sample of the observed magnitudes of \kspn\ during its early phase.
The entire observed magnitudes of \kspn\ are available in the electronic edition.} 
\end{deluxetable} 
\label{tab:mag}

\clearpage
\begin{deluxetable}{lc}
\tablecolumns{2}
\tablewidth{0pt}
\tablecaption{Parameters of \kspn}
\tablehead{ \colhead{Parameter} & \colhead{Value}  }
\startdata
    (RA, decl.) (J2000)                        &   ($\rm 00^h57^m03.19^s, -37\degr02\arcmin23\farcs64$) \\
    First Detection: UT \& MJD                 & 21$\rm h$ 33$\rm m$ 29$\rm s$ on September 24, 2015 \& 57289.89825 \\
    Peak Epoch (\tp)\tablenotemark{$\dagger$}: UT \& MJD & 1$\rm h$ 55$\rm m$ on October 10, 2015 \& 57305.08 \\
    Observed Peak Magnitudes                    &  18.59 ($B$), 18.49 ($V$), 18.91 ($I$) mag  \\ 
    Observed Peak Epochs\tablenotemark{$\ddagger$}            &  --1.11 ($B$), 1.29 ($V$), --2.77 ($I$) days  \\ 
    Redshift ($z$)                             &  0.072 $\pm$ 0.003 \\
    Phillips Parameter (\dm15)                 &  1.62 $\pm$ 0.03 mag \\
    Color Stretch Parameter (\sbv)             &  0.54 $\pm$  0.05  \\
    Absolute Peak Magnitude                   & --18.94 ($B$), --18.93 ($V$), --18.38 ($I$) mag \\
    Early Light Power-law Index  ($\alpha$)    &  2.0 $\pm$ 0.2 ($B$), 1.9 $\pm$ 0.2 ($V$), 2.1 $\pm$ 0.2 ($I$) \\    
    Epoch of First Light ($t_0$)\tablenotemark{$\ddagger$}               & --18.4 $\pm$ 0.6 days    \\
    First Detection from Epoch of First Light  & 3.3 $\pm$ 0.6 days \\
    Peak Bolometric Luminosity \& Epoch  & (9.0 $\pm$ 0.3) $\times$ 10$^{42}$ erg s$^{-1}$ \& --0.5 $\pm$ 0.6 days \\
    \ni56\ mass ($M_{\rm Ni}$)                 & 0.32 $\pm$ 0.01 \msol\ \\
    Ejecta Mass ($M_{\rm ej}$)                 & 0.84 $\pm$ 0.12 \msol\  \\
    Ejecta Kinetic Energy ($E_{\rm K}$)        & (0.61 $\pm$ 0.14) $\times$ 10$^{51}$ erg \\  
\enddata    
\tablenotetext{\dagger}{The epoch of the peak $B$-band brightness determined in the \snp\ template fitting.}
\tablenotetext{\ddagger}{This is relative to \tp.}
\tablecomments{All the date and time values are given in the observer frame.}
\end{deluxetable}
\label{tab:par}

\end{document}